# Simultaneous comparison of the predictive values of two binary diagnostic tests in the presence of categorical covariates

José A. Roldán-Nofuentes[1] and Saad B. Regad[2]

[1]Department of Statistics, School of Medicine, University of Granada, 1816, Granada, Spain

Email: jaroldan@ugr.es. ORCID: https://orcid.org/0000-0003-0251-5588

[2]Epidemiology and Public Health Research Unit and URMCD, School of Medicine, University of Nouakchott Alaasriya, Nouakchott BP 880, Mauritania

Article accepted for publication in the Journal of Biopharmaceutical Statistics

**Abstract.** Comparison of predictive values of diagnostic tests is a topic of interest in Medical Statistics, and has been the subject of different studies. In clinical practice, it is frequent to observe categorical covariates when comparing diagnostic tests. In this framework, a global hypothesis test is proposed to simultaneously compare the predictive values of two diagnostic tests when in all of the individuals categorical covariates are observed. This hypothesis test is solved through regression models and also by weighted least squares method for the analysis of categorical data. Simulation experiments were carried out to study the asymptotic behavior of these methods when a binary covariate is observed and when a covariate with three categories is observed, and these were compared to the behavior of the global test when the covariate is ignored. In general, the method based on regression models has shown to have better asymptotic behavior than the other methods. Furthermore, we studied the application of the method based on the regression models when no covariate is observed, for which the individuals in the sample are randomly assigned to a binary dummy random variable. Simulation experiments carried out showed that this method has greater power than the method without the covariate. The results were applied to two examples.

## 1. Introduction

The use of diagnostic tests is fundamental in clinical and epidemiological practice. In the case of binary diagnostic tests (*BDTs*), sensitivity and specificity are the basic parameters of diagnostic accuracy [1]. Sensitivity (*Se*) is the probability that the *BDT* result is positive when the disease is present, and specificity (*Sp*) is the probability that the *BDT* result is negative when the disease is absent. The evaluation or comparison of the accuracy of *BDTs* is carried out with respect to a gold standard. A gold standard (*GS*) is a medical test that determines without error whether a patient has the disease or not, for example a biopsy. However, in clinical practice, the most important measures for evaluating and comparing *BDTs* are the positive and negative predictive values (*PVs*). The positive *PV* ( $\tau$ ) is the probability of having the disease given a positive *BDT* result, and the negative *PV* ( $\upsilon$ ) is the probability of not having the disease given a negative *BDT* result. Applying the Bayes theorem, the positive *PV* and negative *PV* are expressed as

$$\tau = \frac{p \times Se}{p \times Se + (1-p) \times (1-Sp)} \text{ and } \upsilon = \frac{(1-p) \times Sp}{p \times (1-Se) + (1-p) \times Sp},$$

respectively, where $p$ is the disease prevalence. The most commonly used sample design in clinical practice to compare two *BDTs* is the paired design [1, 2], which consists of applying the two *BDTs* to a sample of $n$ individuals whose disease status (present or absent) is known by applying a gold standard (*GS*). Under a paired design, different works have studied the two-tailed hypothesis tests

$$H_0 : \tau_1 = \tau_2 \text{ vs } H_1 : \tau_1 \neq \tau_2$$

and

$$H_0 : \upsilon_1 = \upsilon_2 \text{ vs } H_1 : \upsilon_1 \neq \upsilon_2 ,$$

that is, the comparisons of the two positive *PVs* and the two negative *PVs* individually.

Bennett [3, 4] has applied the chi-square test to solve each hypothesis test. Jamart [5] has shown that the method proposed by Bennett is not suitable for solving these problems. Leisenring et al [6] have solved the two individual tests by applying *GEE* models, i.e. $\log\left\{P\left(D_{ij}=1\middle|Z_{ij},T_{ij}=1\right)\right\}=\alpha_{\tau}+\beta_{\tau}Z_{ij}$ to compare the positive *PVs* and $\log\left\{P\left(D_{ij}=1\middle|Z_{ij},T_{ij}=1\right)\right\}=\alpha_{\upsilon}+\beta_{\upsilon}Z_{ij}$, where $D_{ij}$ models the disease status (1 for diseased and 0 for non-diseased) for *i*-th individual to whom the *j*-th *BDT* has been applied ($j=1,2$), $Z_{ij}$ indicates the *BDT* applied (1 for *BDT* 1 and 0 for *BDT* 2) and $T_{ij}$ is the result (1 if the result is positive and 0 if it is negative) of the *j*-th *BDT* in the *i*-th individual. The hypothesis test of equality of the two positive (negative) *PVs* is equivalent to the test $H_0:\beta_{\tau}=0$ ($H_0:\beta_{\upsilon}=0$).

Wang et al [7] have solved the two individual hypothesis tests by applying the weighted least squares method for categorical data, proposing a Wald-type test statistic to compare the positive (negative) *PVs* based on the multinomial distribution. The Wald test statistic to compare the two positive *PVs* is $\left(\hat{\tau}_1-\hat{\tau}_2\right)^2/\hat{V}ar\left(\hat{\tau}_1-\hat{\tau}_2\right)$, whose distribution is approximately a chi-square with one degree of freedom, and where the estimators of the positive *PVs* are obtained by means of the classic expressions based on the multinomial distribution and the expression of the variance is obtained by applying the delta method. The Wald test statistic to compare the two negative *PVs* has a similar expression. Wang et al [7] have demonstrated through simulation experiments that their method performs better, in terms of type I error rate and power, than the method of Leisenring et al [6]. The method of Leisenring et al [6] is based on *GEE* models and the method of Wang et al [7] is based on the multinomial distribution, and in both cases the test statistics have complex expressions. Kosinski [8] has derived a weighted generalized score test statistic for comparing the two positive (negative) *PVs*. Each test statistic incorporates an empirical covariance matrix with newly proposed weights, whose expressions are simple and easy to compute and which have approximately a chi-square distribution with one degree of freedom. Kosinski [8] has demonstrated through simulation experiments that his method has a better performance, in terms of type I error rate and power, than the methods of Leisenring et al [6] and Wang et al [7].

Tsou [9] has derived a robust score test statistic to solve each individual test and has shown that his method is identical to the Kosisnki method [8]. Wu [10] has reformulated the test statistic of the method of Bennett [3, 4] to solve each individual hypothesis test and found that the reformulated statistical test is asymptotically equivalent to the method of Wang et al [7]. Takahashi and Yamamoto [11] have deduced an exact test to solve each individual test and have verified through simulations that the exact test controls the type I error rate and has a power similar to other approximate methods. Roldán-Nofuentes and Regad [12] have studied the individual comparison of the *PVs* through confidence intervals for the difference (ratio) of the two positive (negative) *PVs*.

The positive *PV* and negative *PV* of each *BDT* depend on the same parameters (sensitivity and specificity of each *BDT*, and disease prevalence), so the simultaneous comparison of the *PVs* is of interest. This simultaneous comparison consists of solving the global hypothesis test

$$\begin{aligned} H_0 &: \left(\tau_1 = \tau_2 \text{ and } \upsilon_1 = \upsilon_2\right) \\ H_1 &: \left(\tau_1 \neq \tau_2 \text{ and/or } \upsilon_1 \neq \upsilon_2\right). \end{aligned} \quad (1)$$

Moskowitz and Pepe [13] have studied the joint comparison of the positive (negative) *PVs* of two *BDTs* through confidence regions based on the chi-square distribution. Moskowitz and Pepe [13] have proposed the regression model $\log\left(P\left(D=1|W,T_1\neq T_2\right)\right)=\gamma_0+\gamma_1 W$, where $D$ is the variable that models the result of the gold standard (1 when the individual is the diseased and 0 when the individual is not diseased), $T_i$ models the result of $i$-th *BDT* (1 if the result is positive and 0 if it is negative), $W=0$ if $T_1=1$ and $T_2=0$, and $W=1$ if $T_1=0$ and $T_2=1$. In the proposed mdoel, hypothesis test $H_0:\gamma_1=0$ is equivalent to proving that $P\left(D=1|T_1=1,T_2=0\right)=P\left(D=1|T_1=0,T_2=1\right)$. Moskowitz and Pepe [13] have shown that when $P\left(T_1=1\right)=P\left(T_2=1\right)$, the hypothesis test $H_0:\gamma_1=0$ is equivalent to showing that the two positive (negative) *PVs* are equal.

Roldán-Nofuentes et al [14] have deduced a test statistic based on the chi-square distribution to solve the global hypothesis test (1) and have proposed a method based on multiple comparisons to investigate the causes of significance if the hypothesis test is significant at an $\alpha$ error. Roldán-Nofuentes et al [14] have shown through simulation experiments that individual comparison of *PVs* can lead to erroneous conclusions. These simulation experiments have shown that when solving the two individual hypothesis tests, the type I error rate is much higher than the nominal error set, while the statistical test of the global hypothesis test has a good behavior in terms of type I error rate. Roldán-Nofuentes et al [14] illustrated their method with an example of diagnosis of the coronary artery disease using two diagnostic tests, dobutamine echocardiography and myocardial perfusion scintigraphy. The resolution of the two individual tests using the method of Leisenring et al [6] (Wang et al [7]) shows that there are no significant differences between the two positive *PVs*, and that the negative *PV* of dobutamine echocardiography is significantly higher than the negative *PV* of myocardial perfusion scintigraphy. However, the global hypothesis test concludes that there are no significant differences between the *PVs* of both diagnostic tests. Therefore, Roldán-Nofuentes et al have shown that the comparison of *PVs* should be carried out jointly, solving the hypothesis test (1).

Wu [15] has proposed a statistic based on the extension of McNemar's test for the global hypothesis test (1) and has demonstrated through simulations that his method shows an asymptotic behaviour which is as good as or better than the method of Roldán-Nofuentes et al [14].

All of the above methods compare *PVs* under a paired design and consider only the results of the *BDTs*. However, in practice, when assessing or comparing *BDTs*, it is common to observe covariates in all of the individuals. These covariates may be of any type (categorical or quantitative), although the quantitative covariates are normally categorized due to the fact that the analysis of *BDTs* leads to the analysis of contingency tables. In this situation, if the covariate is quantitative, a multitude of tables (one for each value of the covariate) are obtained, which makes the study considerably more difficult. Therefore, from now on it is assumed that the covariates are categorical or have been transformed into categorical ones. In this situation, the comparison of the *PVs* of two *BDTs* has never been studied. For example, in the diagnosis of colorectal cancer, diabetes is a covariate that increases the risk of suffering from this disease, so it is a covariate that must be considered in studies. Thus, if the *PVs* of two *BDTs* are compared for the diagnosis of this cancer, it is important to consider whether the individual is diabetic or not. If any of the previous methods are applied, this covariate would not be taken into account. This is the motivation of our article, to study the simultaneous comparison of the *PVs* of two *BDTs* when categorical covariates are observed in all of the individuals in the sample.

The rest of the manuscript is structured as follows. In Section 2, we propose a method based on regression models to solve the global hypothesis test (1) when categorical covariates are observed in all of the individuals in the sample. In Section 3, the same problem is solved through the weighted least squares method for the analysis of categorical data. In Section 4, simulation experiments are carried out to study the asymptotic behaviour of the methods proposed in different covariate scenarios, and they are compared with the Wu method [15] when the covariate is not used. In Section 5, two functions in *R* [16] are presented to solve the problems studied when a binary covariate is observed and when no covariate is observed and individuals are randomly assigned to a dummy binary random variable. In Section 6, the results are applied to two examples, and in Section 7 the results obtained are discussed.

## 2. Approximation through regression models

Let us consider that two *BDTs* (*Test* 1 and *Test* 2) are applied to all of the individuals in a random sample sized *n*, and their disease status is known through the application of a *GS*. Let us also consider that *C* categorical covariates are observed in all of the individuals. Let $D$ be the random binary variable that models the result of the *GS*, in such a way that $D=0$ when the individual does not have the disease and $D=1$ when the individual does. Let $T_1$ and $T_2$ be the random binary variables that model the result of *Test* 1 and *Test* 2 respectively, in such a way that $T_t=0$ when the result of the test is negative and $T_t=1$ when the result is positive, with $t=1,2$. Let $X_c$ be the random variable that models the result of the *c*-th covariate and $\mathbf{X}=\left(X_1,...,X_C\right)^T$ the vector formed by all of the covariates. Let us consider that there are *M* different covariate patterns and that $\mathbf{X}=\mathbf{x}_m$ represents the $m$-th pattern of covariates. For the $m$-th pattern we obtain the frequencies given in Table 1, where $s_{jkm}$ is the number of individuals in which $D=1$, $T_1=j$, $T_2=k$ and $\mathbf{X}=\mathbf{x}_m$, $r_{jkm}$ is the number of individuals in which $D=0$, $T_1=j$, $T_2=k$ and $\mathbf{X}=\mathbf{x}_m$. This frequency table for $\mathbf{X}=\mathbf{x}_m$ is similar to that obtained under a paired design [6-8]. The total number of individuals that verify $\mathbf{X}=\mathbf{x}_m$ is $n_m=\sum_{j,k=0}^{1}\left(s_{jkm}+r_{jkm}\right)$. Let $D_i$, $T_{1i}$, $T_{2i}$ and $\mathbf{X}_i$ be the values of the previous variables in the $i$-th individual. The overall *PVs* of the $t$-th *Test* are $\tau_t=P\left(D=1|T_t=1\right)$ and $\upsilon_t=P\left(D=0|T_t=0\right)$ respectively, with $t=1,2$.

Table 1. Observed frequencies for $\mathbf{X}=\mathbf{x}_m$.

| | Disease ($D=1$) | | No disease ($D=0$) | |
|---|---|---|---|---|
| | Test 2 result | | Test 2 result | |
| Test 1 result | Positive ($T_2=1$) | Negative ($T_2=0$) | Positive ($T_2=1$) | Negative ($T_2=0$) |
| Positive ($T_1=1$) | $s_{11m}$ | $s_{10m}$ | $r_{11m}$ | $r_{10m}$ |
| Negative ($T_1=0$) | $s_{01m}$ | $s_{00m}$ | $r_{01m}$ | $r_{00m}$ |

Let the probabilities be

$$\phi_{jkm}=P\left(D=1|T_1=j,T_2=k,\mathbf{X}=\mathbf{x}_m\right), \tag{2}$$

$$\varphi_{jkm}=P\left(T_1=j,T_2=k|\mathbf{X}=\mathbf{x}_m\right) \tag{3}$$

and

$$\gamma_m=P\left(\mathbf{X}=\mathbf{x}_m\right), \tag{4}$$

where $\mathbf{x}_m=\left(x_{1m},...,x_{Cm}\right)^T$ is the value of the $m$-th pattern of covariates. The overall *PVs* of *Test* 1 are written in terms of the previous parameters as

$$\tau_1 = \frac{\sum_{k=0}^{1}\sum_{m=1}^{M}\phi_{1km}\varphi_{1km}\gamma_m}{\sum_{k=0}^{1}\sum_{m=1}^{M}\varphi_{1km}\gamma_m} \text{ and } \upsilon_1 = \frac{\sum_{k=0}^{1}\sum_{m=1}^{M}\left(1-\phi_{0km}\right)\varphi_{0km}\gamma_m}{\sum_{k=0}^{1}\sum_{m=1}^{M}\varphi_{0km}\gamma_m}, \tag{5}$$

and those of *Test* 2 as

$$\tau_2 = \frac{\sum_{j=0}^{1}\sum_{m=1}^{M}\phi_{j1m}\varphi_{j1m}\gamma_m}{\sum_{j=0}^{1}\sum_{m=1}^{M}\varphi_{j1m}\gamma_m} \text{ and } \upsilon_2 = \frac{\sum_{j=0}^{1}\sum_{m=1}^{M}\left(1-\phi_{j0m}\right)\varphi_{j0m}\gamma_m}{\sum_{j=0}^{1}\sum_{m=1}^{M}\varphi_{j0m}\gamma_m}, \tag{6}$$

The $\phi_{ijm}$ is modelled through logistic regression [17], i.e.

$$\phi_{ijm} = \frac{\exp\left(\beta_0 + \beta_1 i + \beta_2 j + \boldsymbol{\beta}^{*T}\mathbf{X}\right)}{1+\exp\left(\beta_0 + \beta_1 i + \beta_2 j + \boldsymbol{\beta}^{*T}\mathbf{X}\right)}, \tag{7}$$

where $\boldsymbol{\beta}^* = \left(\beta_1^*,...,\beta_M^*\right)^T$, $i$ represents the result of *Test* 1 and $j$ represents the result of *Test* 2, with $i,j = 0,1$ and $m = 1,...,M$.

Regarding the parameter $\varphi_{ijm}$, this is modelled through a multinomial logit regression [18]. Considering, without loss of generality, as a category of reference the result $\left(T_1 = 1, T_2 = 1\right) = \left(i,j\right) = \left(1,1\right)$, then the model is

$$\begin{aligned} \varphi_{00m} &= \frac{\exp\left(\alpha_{00m} + \boldsymbol{\alpha}_{00m}^T\mathbf{x}_m\right)}{1+\sum_{\substack{r,s=0\\(r,s)\neq(1,1)}}^{1}\exp\left(\alpha_{rsm} + \boldsymbol{\alpha}_{rsm}^T\mathbf{x}_m\right)}, \quad \varphi_{01m} = \frac{\exp\left(\alpha_{01m} + \boldsymbol{\alpha}_{01m}^T\mathbf{x}_m\right)}{1+\sum_{\substack{r,s=0\\(r,s)\neq(1,1)}}^{1}\exp\left(\alpha_{rsm} + \boldsymbol{\alpha}_{rsm}^T\mathbf{x}_m\right)}, \\ \varphi_{10m} &= \frac{\exp\left(\alpha_{10m} + \boldsymbol{\alpha}_{10m}^T\mathbf{x}_m\right)}{1+\sum_{\substack{r,s=0\\(r,s)\neq(1,1)}}^{1}\exp\left(\alpha_{rsm} + \boldsymbol{\alpha}_{rsm}^T\mathbf{x}_m\right)} \text{ and } \varphi_{11m} = \frac{1}{1+\sum_{\substack{r,s=0\\(r,s)\neq(1,1)}}^{1}\exp\left(\alpha_{rsm} + \boldsymbol{\alpha}_{rsm}^T\mathbf{x}_m\right)}, \end{aligned} \tag{8}$$

where $\boldsymbol{\alpha}_{ijm} = \left(\alpha_{ijm1},...,\alpha_{ijmK}\right)^T$, with $\left(i,j\right) \neq \left(1,1\right)$.

Finally, regarding the probability of each covariate pattern, $\gamma_m = P(\mathbf{X} = \mathbf{x}_m)$, this is the probability of a cell of a multinomial distribution, and it is verified that $\sum_{m=1}^{M} \gamma_m = 1$.

Let the vectors be $\boldsymbol{\beta} = \left(\beta_0, \beta_1, \beta_2, \boldsymbol{\beta}^{*T}\right)^T$, $\boldsymbol{\alpha} = \left(\alpha_{001}, \alpha_{011}, ..., \boldsymbol{\alpha}_{01M}^T, \boldsymbol{\alpha}_{10M}^T\right)^T$ and $\boldsymbol{\gamma} = \left(\gamma_1, ..., \gamma_M\right)^T$. The log-likelihood function based on the observed data is

$$\begin{gathered} l(\boldsymbol{\alpha}, \boldsymbol{\beta}, \boldsymbol{\gamma}) = \sum_{i=1}^{n} \log\left[P\left(D_i, T_{1i}, T_{2i}, \mathbf{X}_i\right)\right] = \\ \sum_{i=1}^{n} \log\left[P\left(D_i | T_{1i}, T_{2i}, \mathbf{X}_i\right) P\left(T_{1i}, T_{2i} | \mathbf{X}_i\right) P\left(\mathbf{X}_i\right)\right] = \\ \sum_{j,k=0}^{1} \sum_{m=1}^{M} \left(s_{jkm} + r_{jkm}\right) \log\left(\varphi_{jkm}\right) + \sum_{j,k=0}^{1} \sum_{m=1}^{M} s_{jkm} \log\left(\phi_{jkm}\right) + \\ \sum_{j,k=0}^{1} \sum_{m=1}^{M} r_{jkm} \log\left(1 - \phi_{jkm}\right) + \sum_{m=1}^{M} n_m \log\left(\gamma_m\right) = \\ l(\boldsymbol{\alpha}) + l(\boldsymbol{\beta}) + l(\boldsymbol{\gamma}), \end{gathered} \tag{9}$$

where

$$l(\boldsymbol{\alpha}) = \sum_{j,k=0}^{1} \sum_{m=1}^{M} \left(s_{jkm} + r_{jkm}\right) \log\left(\varphi_{jkm}\right), \tag{10}$$

$$l(\boldsymbol{\beta}) = \sum_{j,k=0}^{1} \sum_{m=1}^{M} s_{jkm} \log\left(\phi_{jkm}\right) + \sum_{j,k=0}^{1} \sum_{m=1}^{M} r_{jkm} \log\left(1 - \phi_{jkm}\right) \tag{11}$$

and

$$l(\boldsymbol{\gamma}) = \sum_{m=1}^{M} n_m \log\left(\gamma_m\right). \tag{12}$$

Consequently, the *MLEs* of the parameters $\boldsymbol{\alpha}$, $\boldsymbol{\beta}$ and $\boldsymbol{\gamma}$ are obtained maximizing the functions (10), (11) and (12) respectively. The vectors $\boldsymbol{\alpha}$ and $\boldsymbol{\beta}$ are estimated using a statistical software, e.g. *R* [16], as well as their corresponding variance-covariance matrixes. Regarding the vector $\boldsymbol{\gamma}$, the *MLE* is $\hat{\gamma}_m = n_m / n$ and its estimated variance-covariance matrix is

$$\hat{\Sigma}_{\hat{\boldsymbol{\gamma}}} = \frac{Diag(\hat{\boldsymbol{\gamma}}) - \hat{\boldsymbol{\gamma}}\hat{\boldsymbol{\gamma}}^T}{n} \tag{13}$$

Let $\boldsymbol{\theta} = (\tau_1, \upsilon_1, \tau_2, \upsilon_2)^T$ be the vector formed by the overall *PVs*. As the overall *PVs* are functions of $\boldsymbol{\alpha}$, $\boldsymbol{\beta}$ and $\boldsymbol{\gamma}$, the variance-covariance matrix of *PVs* is estimated applying the delta method [19], i.e. the estimated variance-covariance matrix of $\hat{\boldsymbol{\theta}}$ is

$$\hat{\Sigma}_{\hat{\boldsymbol{\theta}}} = \left(\frac{\partial \boldsymbol{\theta}}{\partial \boldsymbol{\alpha}}\right)_{\boldsymbol{\alpha}=\hat{\boldsymbol{\alpha}}} \hat{\Sigma}_{\hat{\boldsymbol{\alpha}}} \left(\frac{\partial \boldsymbol{\theta}}{\partial \boldsymbol{\alpha}}\right)^T_{\boldsymbol{\alpha}=\hat{\boldsymbol{\alpha}}} + \left(\frac{\partial \boldsymbol{\theta}}{\partial \boldsymbol{\beta}}\right)_{\boldsymbol{\beta}=\hat{\boldsymbol{\beta}}} \hat{\Sigma}_{\hat{\boldsymbol{\beta}}} \left(\frac{\partial \boldsymbol{\theta}}{\partial \boldsymbol{\beta}}\right)^T_{\boldsymbol{\beta}=\hat{\boldsymbol{\beta}}} + \left(\frac{\partial \boldsymbol{\theta}}{\partial \boldsymbol{\gamma}}\right)_{\boldsymbol{\gamma}=\hat{\boldsymbol{\gamma}}} \hat{\Sigma}_{\hat{\boldsymbol{\gamma}}} \left(\frac{\partial \boldsymbol{\theta}}{\partial \boldsymbol{\gamma}}\right)^T_{\boldsymbol{\gamma}=\hat{\boldsymbol{\gamma}}}.$$

The global hypothesis test (1) to check the equality of the overall *PVs* is equivalent to the hypothesis test

$$H_0 : \boldsymbol{\psi\theta} = \mathbf{0} \text{ vs } H_1 : \boldsymbol{\psi\theta} \neq \mathbf{0}.$$

where $\boldsymbol{\psi}$ is a complete range matrix sized $2 \times 4$ whose elements are known constants, i.e.

$$\boldsymbol{\psi} = \begin{pmatrix} 1 & 0 & -1 & 0 \\ 0 & 1 & 0 & -1 \end{pmatrix}.$$

Applying the multivariate central limit theorem it is verified that

$$\sqrt{n}\left(\hat{\boldsymbol{\theta}} - \boldsymbol{\theta}\right) \xrightarrow[n \to \infty]{} N_4\left(\mathbf{0}, \boldsymbol{\Sigma}_{\boldsymbol{\theta}}\right).$$

Then, the test statistic

$$Q = \hat{\boldsymbol{\theta}}^T \boldsymbol{\psi}^T \left(\boldsymbol{\psi} \hat{\Sigma}_{\hat{\boldsymbol{\theta}}} \boldsymbol{\psi}^T\right)^{-1} \boldsymbol{\psi}\hat{\boldsymbol{\theta}}$$

is distributed according to Hotelling's *T*-squared distribution with a dimension of 2 and *n* degrees of freedom, where 2 is the dimension of the vector $\boldsymbol{\psi}\hat{\boldsymbol{\theta}}$. When *n* is large, the test statistic $Q$ is distributed according to a central chi-squared distribution with 2 degrees of freedom when the null hypothesis is true, i.e.

$$Q = \hat{\boldsymbol{\theta}}^T \boldsymbol{\psi}^T \left( \boldsymbol{\psi} \hat{\Sigma}_{\hat{\boldsymbol{\theta}}} \boldsymbol{\psi}^T \right)^{-1} \boldsymbol{\psi} \hat{\boldsymbol{\theta}} \xrightarrow[n \to \infty]{} \chi_2^2 . \tag{14}$$

Regarding individual comparisons, $H_0 : \tau_1 = \tau_2$ vs $H_1 : \tau_1 \neq \tau_2$ and $H_0 : \upsilon_1 = \upsilon_2$ vs $H_1 : \upsilon_1 \neq \upsilon_2$, the test statistic to compare the two overall positive *PVs* is

$$z = \frac{\hat{\tau}_1 - \hat{\tau}_2}{\sqrt{\hat{V}ar(\hat{\tau}_1) + \hat{V}ar(\hat{\tau}_2) - 2\hat{C}ov(\hat{\tau}_1, \hat{\tau}_2)}} \xrightarrow[n \to \infty]{} N(0,1), \tag{15}$$

and the test statistic to compare the two overall negative *PVs* is

$$z = \frac{\hat{\upsilon}_1 - \hat{\upsilon}_2}{\sqrt{\hat{V}ar(\hat{\upsilon}_1) + \hat{V}ar(\hat{\upsilon}_2) - 2\hat{C}ov(\hat{\upsilon}_1, \hat{\upsilon}_2)}} \xrightarrow[n \to \infty]{} N(0,1). \tag{16}$$

When the global test (1) is significant to set $\alpha$, the causes of the significance are investigated in a similar way to the method proposed by Roldán-Nofuentes el al [14], i.e. the two individual hypothesis tests are solved and the *p*-values obtained are adjusted through a method of multiple comparisons e.g. the Holm method [20].

## 3. Approximation using weighted least squares

In this section, the problem is approximated using weighted least squares (*WLS*) for analysis of categorical data [19]. To simplify, we are going to consider that in the case of all individuals we observe a categorical covariate with *M* different patterns. Using the same notation as in the previous section (equations (2), (3) and (4)), the log-likelihood function based on the observed data is

$$
\begin{aligned}
l(\boldsymbol{\phi},\boldsymbol{\varphi},\boldsymbol{\gamma}) &= \sum_{i=1}^{n}\log\left[P\left(D_i,T_{1i},T_{2i},\mathbf{X}_i\right)\right]= \\
&\sum_{i=1}^{n}\log\left[P\left(D_i\left|T_{1i},T_{2i},\mathbf{X}_i\right.\right)P\left(T_{1i},T_{2i}\left|\mathbf{X}_i\right.\right)P\left(\mathbf{X}_i\right)\right]= \\
&\sum_{j,k=0}^{1}\sum_{m=1}^{M}s_{jkm}\log\left(\phi_{jkm}\right)+\sum_{j,k=0}^{1}\sum_{m=1}^{M}r_{jkm}\log\left(1-\phi_{jkm}\right)+ \\
&\sum_{j,k=0}^{1}\sum_{m=1}^{M}\left(s_{jkm}+r_{jkm}\right)\log\left(\varphi_{jkm}\right)+\sum_{m=1}^{M}n_m\log\left(\gamma_m\right)=l(\boldsymbol{\phi})+l(\boldsymbol{\varphi})+l(\boldsymbol{\gamma}).
\end{aligned}
\tag{17}
$$

The parameters $\boldsymbol{\phi}$ and $\boldsymbol{\varphi}$ are not modeled by the regression models given in the previous Section, and their estimators are obtained by maximizing the functions $l(\boldsymbol{\phi})$ and $l(\boldsymbol{\varphi})$ respectively. The parameters $\boldsymbol{\phi}$, $\boldsymbol{\varphi}$ and $\boldsymbol{\gamma}$ are different and the functions $l(\boldsymbol{\phi})$, $l(\boldsymbol{\varphi})$ and $l(\boldsymbol{\gamma})$ are log-likelihood functions of multinomial distributions, then maximizing each of these functions we obtain that the *MLE*s of these parameters are

$$
\hat{\phi}_{jkm}^{*}=\frac{s_{jkm}}{s_{jkm}+r_{jkm}},\ \hat{\varphi}_{jkm}^{*}=\frac{s_{jkm}+r_{jkm}}{n_m}\ \text{and}\ \hat{\gamma}_m=\frac{n_m}{n}. \tag{18}
$$

Substituting in the expressions of the overall *PVs* each parameter with its *MLE*, it is obtained that the *MLEs* values of the overall *PVs* of *Test* 1 are

$$
\hat{\tau}_1^{*}=\frac{\sum_{k=0}^{1}\sum_{m=1}^{M}s_{1km}}{\sum_{k=0}^{1}\sum_{m=1}^{M}n_{1km}}\ \text{and}\ \hat{\upsilon}_1^{*}=\frac{\sum_{k=0}^{1}\sum_{m=1}^{M}r_{0km}}{\sum_{k=0}^{1}\sum_{m=1}^{M}n_{0km}}, \tag{19}
$$

and those of *Test* 2 are

$$
\hat{\tau}_2^{*}=\frac{\sum_{j=0}^{1}\sum_{m=1}^{M}s_{j1m}}{\sum_{j=0}^{1}\sum_{m=1}^{M}n_{j1m}}\ \text{and}\ \hat{\upsilon}_2^{*}=\frac{\sum_{j=0}^{1}\sum_{m=1}^{M}r_{j0m}}{\sum_{j=0}^{1}\sum_{m=1}^{M}n_{j0m}}, \tag{20}
$$

where $n_{jkm}=s_{jkm}+r_{jkm}$. Regarding the variance-covariance matrix of the overall estimators, applying the delta method [19] it is obtained that

$$\hat{\Sigma}_{\hat{\boldsymbol{\theta}}}^{*} = \left(\frac{\partial \boldsymbol{\theta}}{\partial \boldsymbol{\phi}}\right)_{\boldsymbol{\phi}=\hat{\boldsymbol{\phi}}^{*}} \hat{\Sigma}_{\hat{\boldsymbol{\phi}}} \left(\frac{\partial \boldsymbol{\theta}}{\partial \boldsymbol{\phi}}\right)_{\boldsymbol{\phi}=\hat{\boldsymbol{\phi}}^{*}}^{T} + \left(\frac{\partial \boldsymbol{\theta}}{\partial \boldsymbol{\varphi}}\right)_{\boldsymbol{\varphi}=\hat{\boldsymbol{\varphi}}^{*}} \hat{\Sigma}_{\hat{\boldsymbol{\varphi}}} \left(\frac{\partial \boldsymbol{\theta}}{\partial \boldsymbol{\varphi}}\right)_{\boldsymbol{\varphi}=\hat{\boldsymbol{\varphi}}^{*}}^{T} + \left(\frac{\partial \boldsymbol{\theta}}{\partial \boldsymbol{\gamma}}\right)_{\boldsymbol{\gamma}=\hat{\boldsymbol{\gamma}}} \hat{\Sigma}_{\hat{\boldsymbol{\gamma}}} \left(\frac{\partial \boldsymbol{\theta}}{\partial \boldsymbol{\gamma}}\right)_{\boldsymbol{\gamma}=\hat{\boldsymbol{\gamma}}}^{T}.$$

where

$$\hat{\Sigma}_{\hat{\boldsymbol{\phi}}}^{*} = Diag\left\{\hat{\Sigma}_{\hat{\boldsymbol{\phi}}_1}^{*},...,\hat{\Sigma}_{\hat{\boldsymbol{\phi}}_M}^{*}\right\}$$

and

$$\hat{\Sigma}_{\hat{\boldsymbol{\varphi}}}^{*} = Diag\left\{\hat{\Sigma}_{\hat{\boldsymbol{\varphi}}_1}^{*},...,\hat{\Sigma}_{\hat{\boldsymbol{\varphi}}_M}^{*}\right\}.$$

Matrix $\hat{\Sigma}_{\hat{\boldsymbol{\phi}}_m}$ is [21]

$$\hat{\Sigma}_{\hat{\boldsymbol{\phi}}_m^{*}}^{*} = Diag\left\{\frac{\hat{\phi}_{jkm}^{*2}\left(1-\hat{\phi}_{jkm}^{*}\right)^{2}}{s_{jkm}\left(1-\hat{\phi}_{jkm}^{*}\right)^{2} + r_{jkm}\hat{\phi}_{jkm}^{*2}}\right\}$$

and matrix $\hat{\Sigma}_{\hat{\boldsymbol{\varphi}}_m}$ is

$$\hat{\Sigma}_{\hat{\boldsymbol{\varphi}}_m^{*}}^{*} = \frac{Diag\left(\hat{\boldsymbol{\varphi}}_m^{*}\right) - \hat{\boldsymbol{\varphi}}_m^{*}\hat{\boldsymbol{\varphi}}_m^{*T}}{n_m}$$

with $\hat{\boldsymbol{\phi}}_m^{*} = \left(\hat{\phi}_{11m}^{*},\hat{\phi}_{10m}^{*},\hat{\phi}_{01m}^{*},\hat{\phi}_{00m}^{*}\right)^{T}$, $\hat{\boldsymbol{\varphi}}_m^{*} = \left(\hat{\varphi}_{11m}^{*},\hat{\varphi}_{10m}^{*},\hat{\varphi}_{01m}^{*},\hat{\varphi}_{00m}^{*}\right)^{T}$, $\hat{\boldsymbol{\phi}}^{*} = \left(\hat{\boldsymbol{\phi}}_1^{*},...,\hat{\boldsymbol{\phi}}_M^{*}\right)$ and $\hat{\boldsymbol{\varphi}}^{*} = \left(\hat{\boldsymbol{\varphi}}_1^{*},...,\hat{\boldsymbol{\varphi}}_M^{*}\right)$. Matrix $\hat{\Sigma}_{\hat{\boldsymbol{\gamma}}}$ is the same as the one given in expression (13). The elements of matrix $\hat{\Sigma}_{\hat{\theta}}^{*}$ have long and complicated expressions that require statistical software for their calculation. Finally, the test statistics to solve the hypothesis tests have the same expressions as those given in Section 2, and if the global test is significant, the causes of the significance are investigated in a similar way to in Section 2.

The *MLEs* of the overall *PVs*, expressions (19) and (20), are equivalent to the *MLEs* of the *PVs* without considering the covariate (which are obtained when we add all of the frequencies tables without considering the patterns of the covariate). For example, if the covariate is binary, the overall estimators of the overall *PVs* of *Test* 1 are

$$\hat{\tau}_1^* = \frac{s_{10}+s_{11}}{s_{10}+s_{11}+r_{10}+r_{11}} \text{ and } \hat{\upsilon}_1^* = \frac{r_{00}+r_{01}}{s_{00}+s_{01}+r_{00}+r_{01}}, \tag{21}$$

and those of *Test* 2 are

$$\hat{\tau}_2^* = \frac{s_{01}+s_{11}}{s_{01}+s_{11}+r_{01}+r_{11}} \text{ and } \hat{\upsilon}_2^* = \frac{r_{00}+r_{10}}{s_{00}+s_{10}+r_{00}+r_{10}}, \tag{22}$$

where $s_{jk} = s_{jk1} + s_{jk2}$ and $r_{jk} = r_{jk1} + r_{jk2}$. The *MLEs* of the overall *PVs* coincide with the estimators of the *PVs* without considering the covariate (see Appendix I). However, the estimated variances-covariances are different in both situations. Appendix I shows the expressions of the variances-covariances of the estimators of the *PVs* without considering the covariate.

If for all individuals more than one categorical covariate is observed, the problem is solved in a similar way. Thus, for example, if we observe a covariate with two categories and others with four, the model would be applied considering a covariate with eight categories [17].

## 4. Simulation experiments

Monte Carlo simulation experiments were carried out to study the type I error rates and the powers of the methods proposed in Sections 2 and 3, and the relative biases of the estimators, in three different situations: a binary covariate, a covariate with three categories, and when no covariate is observed and a dummy random variate is generated.

### *4.1. A binary covariate*

This situation is very common in clinical practice, in which the researcher observes a binary covariate in all individuals, for example sex. In this situation, when the *BDTs*, the *GS* and the binary covariate are observed in all of the individuals in a random sample, two frequencies tables are observed. The theoretical probabilities in each table are:

$$\begin{gathered}P\left(D=1,T_1=j,T_2=k,X=m\right)=\\ \gamma_m p_m\left[\frac{\tau_{1m}^{j}\left(\upsilon_{1m}-q_m\right)^{j}\left(\tau_{1m}+p_m\right)^{1-j}\left(1-\upsilon_{1m}\right)^{1-j}}{p_m^{j}p_m^{1-j}Y_{1m}^{j}Y_{1m}^{1-j}}\times\right.\\ \left.\frac{\tau_{2m}^{k}\left(\upsilon_{2m}-q_m\right)^{k}\left(\tau_{2m}+p_m\right)^{1-k}\left(1-\upsilon_{2m}\right)^{1-k}}{p_m^{k}p_m^{1-k}Y_{2m}^{k}Y_{2m}^{1-k}}+\delta_{ik}\varepsilon_{1m}\right],\\ P\left(D=0,T_1=j,T_2=k,X=m\right)=\\ \gamma_m q_m\left[\frac{\left(1-\tau_{1m}\right)^{j}\left(\upsilon_{1m}-q_m\right)^{j}\left(\tau_{1m}-p_m\right)^{1-j}\upsilon_{1m}^{1-j}}{q_m^{j}q_m^{1-j}Y_{1m}^{j}Y_{1m}^{1-j}}\times\right.\\ \left.\frac{\left(1-\tau_{2m}\right)^{k}\left(\upsilon_{2m}-q_m\right)^{k}\left(\tau_{2m}-p_m\right)^{1-k}\upsilon_{2m}^{1-k}}{q_m^{k}q_m^{1-k}Y_{2m}^{k}Y_{2m}^{1-k}}+\delta_{jk}\varepsilon_{0m}\right],\end{gathered}\tag{23}$$

with $j,k=0,1$, $m=1,2$, verifying that

$$\sum_{i,j=0}^{1}\sum_{m=1}^{2}P\left(D=0,T_1=j,T_2=k,X=m\right)+\sum_{i,j=0}^{1}\sum_{m=1}^{2}P\left(D=1,T_1=j,T_2=k,X=m\right)=1$$

and that $\gamma_2=1-\gamma_1$. In probabilities (23) $p_m=P\left(D=1|X=m\right)$ is the disease prevalence when $X=m$, $q_m=1-p_m$, $\tau_{tm}$ and $\upsilon_{tm}$ are the positive *PV* and negative *PV* of the *t*-th Test ($t=1,2$) when $X=m$ respectively, $Y_{1m}=\tau_{1m}+\upsilon_{1m}-1$, $Y_{2m}=\tau_{2m}+\upsilon_{2m}-1$,

$$\varepsilon_{1m}=\frac{\tau_{1m}\tau_{2m}\left(\upsilon_{1m}-q_m\right)\left(\upsilon_{2m}-q_m\right)\left(\alpha_{1m}-1\right)}{p_m^2Y_{1m}Y_{2m}}$$

and

$$\varepsilon_{0m}=\frac{\left(1-\tau_{1m}\right)\left(1-\tau_{2m}\right)\left(\upsilon_{1m}-q_m\right)\left(\upsilon_{2m}-q_m\right)\left(\alpha_{0m}-1\right)}{q_m^2Y_{1m}Y_{2m}},$$

with $\delta_{jk}=1$ if $j=k$ and $\delta_{jk}=-1$ if $j\neq k$. The parameter $\varepsilon_{1m}$ $\left(\varepsilon_{0m}\right)$ is the covariance between the two *BDTs* subject to the additive model [22] when $D=1$ ($D=0$) and $X=m$, and the parameter $\alpha_{1m}$ ($\alpha_{0m}$) is the covariance between the two *BDTs* subject to the multiplicative model [23] when $D=1$ ($D=0$) and $X=m$, verifying that

$$1\leq\alpha_{1m}\leq\frac{1}{\max\left\{\frac{\tau_{1m}\left(\upsilon_{1m}-q_m\right)}{p_mY_{1m}},\frac{\tau_{2m}\left(\upsilon_{2m}-q_m\right)}{p_mY_{2m}}\right\}}$$

and

$$1\leq\alpha_{0m}\leq\frac{1}{\max\left\{\left(1-\frac{\upsilon_{1m}\left(\tau_{1m}-p_m\right)}{q_mY_{1m}}\right),\left(1-\frac{\upsilon_{2m}\left(\tau_{2_m}-p_m\right)}{q_mY_{2m}}\right)\right\}}.$$

If $\alpha_{1m}=\alpha_{0m}=1$ then the two *BDTs* are conditionally independent on the disease when $X=m$, a situation which is not realistic in practice, and therefore it must be verified that $\alpha_{1m}>1$ and/or $\alpha_{0m}>1$.

The simulation experiments consisted of generating $N=10000$ random samples of multinomial distributions sized $n=\{50,100,200,300,400,500\}$, and whose probabilities were calculated from expressions (23) setting the values of the parameters involved. As values of $\gamma_1$ we took the values 0.60 and 0.80, and as values of $p_m$ we considered the values 25% and 75%. For the *PVs*, we took the values $\tau_{tm},\upsilon_{tm}=\{0.75,0.80,...,0.95\}$, and as values of $\alpha_{1m}$ and $\alpha_{0m}$ we considered intermediate and high values, i.e.

$$\alpha_{1m}=\frac{f}{\max\left\{\frac{\tau_{1m}\left(\upsilon_{1m}-q_m\right)}{p_mY_{1m}},\frac{\tau_{2m}\left(\upsilon_{2m}-q_m\right)}{p_mY_{2m}}\right\}}+1-f$$

and

$$\alpha_{0m} = \frac{f}{\max\left\{\left(1-\frac{(\tau_{1m}-p_m)\upsilon_{1m}}{q_m Y_{1m}}\right),\left(1-\frac{(\tau_{2m}-p_m)\upsilon_{2m}}{q_m Y_{2m}}\right)\right\}} + 1 - f,$$

with $f = \{0.50, 0.90\}$. Once we set the values of $\tau_{tm}$, $\upsilon_{tm}$ and of $p_m$, the overall *PVs* are calculated as

$$\tau_t = \frac{\sum_{m=1}^{2} \tau_{tm}\eta_{tm}\gamma_m}{\sum_{m=1}^{2} \eta_{tm}\gamma_m} \text{ and } \upsilon_t = \frac{\sum_{m=1}^{2} \upsilon_{tm}(1-\eta_{tm})\gamma_m}{\sum_{m=1}^{2}(1-\eta_{tm})\gamma_m},$$

where

$$\eta_{tm} = p_m Se_{tm} + q_m (1 - Sp_{tm})$$

and $Se_{tm}$ and $Sp_{tm}$ are the sensitivity and the specificity of the *t*-th *Test* when $X = m$, and whose expressions are

$$Se_{tm} = \frac{\tau_{tm}(\upsilon_{tm} - q_m)}{p_m Y_{tm}} \text{ and } Sp_{tm} = \frac{\upsilon_{tm}(\tau_{tm} - p_m)}{q_m Y_{tm}}.$$

Therefore, the probabilities of the multinomial distributions were calculated from the *PVs*. The simulation experiments were designed in such a way that in all of the samples generated it is possible to estimate the *PVs* and their variances-covariances. If in a sample there are many frequencies equal to zero, then the proposed models cannot be applied, so in this situation that sample was disregarded and another was generated in its place. In each scenario, we calculated the type I error rates or powers, depending on the case, applying the methods in sections 2 and 3 (*RM* and *WLS* respectively) and we also applied the Wu method [15] (*WM*). Regarding the *WM*, this is summarized in Appendix II. We compare our methods proposed here only with the *WM* [15], since the *WM* [15] solves the global test without considering the covariate and we do not apply the method of Roldán-Nofuentes et al [14] since this method does not offer better results than the *WM*.

Likewise, individual methods of comparison of positive (negative) *PVs* are not considered since these do not solve the global hypothesis test. In these simulation experiments, we also calculated the relative biases of the estimators of the overall *PVs* applying the *RM* and *WLS* methods. The simulation experiments were carried out with *R* [16] and as a nominal error we set $\alpha = 5\%$.

Table 2 shows the type I error rates of the *RM*, *WLS* and *WM* for different scenarios. The three methods have a very similar type I error rates, and these three methods are conservative for the sample sizes considered. The probability $\gamma_1$ does not have an important effect on the type I error rates of each method, whereas the increase in the covariances involves a decrease in the type I error rates.

Table 3 shows the powers of the methods for different scenarios. In general terms, the power of the *RM* method is always higher than the powers of the *WLS* and *WM* methods, above all when the size of the sample is moderate ($n = 100 - 200$) or the sample size is large ($n \geq 300$). In some scenarios the power of the *RM* method may be more than 10% higher than the power of the *WM* when the sample size is moderate. The increase in the probability $\gamma_1$ involves an increase in the power of the method, above all when the sample size is small ($n = 50$) or moderate ($n = 100 - 200$); the increase in $\gamma_1$ has practically no effect on the power of each method when the sample size is large $(n \geq 300)$. The increase in the covariances involves an increase in the power of each method, above all when the sample size is large.

From the above results, it can be concluded that the method based on regression models has a good asymptotic behavior. Although it is a conservative test (like the other two methods), its power is reasonably good with a moderate sample size.

Table 2. Type I error rates (in %) of the global hypothesis test with a binary covariate.

| | $\tau_{11}=\tau_{21}=\tau_{12}=\tau_{22}=0.80$ $\upsilon_{11}=\upsilon_{21}=\upsilon_{12}=\upsilon_{22}=0.85$ $\tau_1=\tau_2=0.80$ $\upsilon_1=\upsilon_2=0.85$ | | | | | | | | | | | |
|---|---|---|---|---|---|---|---|---|---|---|---|---|
| | $\gamma_1=0.60$ $p_1=75\%$ $p_2=25\%$ | | | | | | $\gamma_1=0.80$ $p_1=75\%$ $p_2=25\%$ | | | | | |
| | $\alpha_{11}=1.00$ $\alpha_{01}=1.18$ $\alpha_{12}=1.52$ $\alpha_{02}=12.69$ | | | $\alpha_{11}=1.01$ $\alpha_{01}=1.32$ $\alpha_{12}=1.93$ $\alpha_{02}=22.04$ | | | $\alpha_{11}=1.00$ $\alpha_{01}=1.18$ $\alpha_{12}=1.52$ $\alpha_{02}=12.69$ | | | $\alpha_{11}=1.01$ $\alpha_{01}=1.32$ $\alpha_{12}=1.93$ $\alpha_{02}=22.04$ | | |
| *n* | *RM* | *WLS* | *WM* | *RM* | *WLS* | *WM* | *RM* | *WLS* | *WM* | *RM* | *WLS* | *WM* |
| 50 | 0 | 0 | 0 | 0 | 0 | 0 | 0 | 0 | 0 | 0 | 0 | 0 |
| 100 | 0.0 | 0.0 | 0.0 | 0 | 0 | 0 | 0.0 | 0.0 | 0.0 | 0 | 0 | 0 |
| 200 | 0.5 | 0.4 | 0.4 | 0 | 0 | 0 | 0.5 | 0.4 | 0.3 | 0 | 0 | 0 |
| 300 | 1.3 | 1.3 | 1.2 | 0 | 0 | 0 | 1.2 | 1.1 | 1.0 | 0 | 0 | 0 |
| 400 | 2.0 | 2.1 | 2.0 | 0 | 0 | 0 | 1.6 | 1.5 | 1.5 | 0.0 | 0.0 | 0.0 |
| 500 | 2.5 | 2.6 | 2.5 | 0.0 | 0.0 | 0.0 | 1.8 | 1.8 | 1.6 | 0.0 | 0.0 | 0.0 |
| | $\gamma_1=0.60$ $p_1=25\%$ $p_2=75\%$ | | | | | | $\gamma_1=0.80$ $p_1=25\%$ $p_2=75\%$ | | | | | |
| | $\alpha_{11}=1.51$ $\alpha_{01}=12.69$ $\alpha_{12}=1.00$ $\alpha_{02}=1.18$ | | | $\alpha_{11}=1.93$ $\alpha_{01}=22.04$ $\alpha_{12}=1.01$ $\alpha_{02}=1.32$ | | | $\alpha_{11}=1.51$ $\alpha_{01}=12.69$ $\alpha_{12}=1.00$ $\alpha_{02}=1.18$ | | | $\alpha_{11}=1.93$ $\alpha_{01}=22.04$ $\alpha_{12}=1.01$ $\alpha_{02}=1.32$ | | |
| *n* | *RM* | *WLS* | *WM* | *RM* | *WLS* | *WM* | *RM* | *WLS* | *WM* | *RM* | *WLS* | *WM* |
| 50 | 0 | 0 | 0 | 0 | 0 | 0 | 0 | 0 | 0 | 0 | 0 | 0 |
| 100 | 0.1 | 0.0 | 0.0 | 0 | 0 | 0 | 0.1 | 0.0 | 0.0 | 0 | 0 | 0 |
| 200 | 0.8 | 0.6 | 0.4 | 0 | 0 | 0 | 0.8 | 0.7 | 0.7 | 0 | 0 | 0 |
| 300 | 1.5 | 1.4 | 1.3 | 0.0 | 0 | 0 | 1.8 | 1.5 | 1.4 | 0.0 | 0 | 0 |
| 400 | 2.1 | 1.9 | 1.8 | 0.0 | 0.0 | 0.0 | 2.3 | 2.2 | 2.1 | 0.1 | 0 | 0 |
| 500 | 2.6 | 2.5 | 2.4 | 0.0 | 0.0 | 0.0 | 2.5 | 2.3 | 2.2 | 0.1 | 0.0 | 0.0 |
| | $\tau_{11}=\tau_{21}=\tau_{12}=\tau_{22}=0.90$ $\upsilon_{11}=\upsilon_{21}=\upsilon_{12}=\upsilon_{22}=0.80$ $\tau_1=\tau_2=0.90$ $\upsilon_1=\upsilon_2=0.80$ | | | | | | | | | | | |
| | $\gamma_1=0.60$ $p_1=75\%$ $p_2=25\%$ | | | | | | $\gamma_1=0.80$ $p_1=75\%$ $p_2=25\%$ | | | | | |
| | $\alpha_{11}=1.03$ $\alpha_{01}=2.09$ $\alpha_{12}=2.44$ $\alpha_{02}=53$ | | | $\alpha_{11}=1.05$ $\alpha_{01}=2.96$ $\alpha_{12}=3.6$ $\alpha_{02}=94.6$ | | | $\alpha_{11}=1.03$ $\alpha_{01}=2.09$ $\alpha_{12}=2.44$ $\alpha_{02}=53$ | | | $\alpha_{11}=1.05$ $\alpha_{01}=2.96$ $\alpha_{12}=3.6$ $\alpha_{02}=94.6$ | | |
| *n* | *RM* | *WLS* | *WM* | *RM* | *WLS* | *WM* | *RM* | *WLS* | *WM* | *RM* | *WLS* | *WM* |
| 50 | 0 | 0 | 0 | 0 | 0 | 0 | 0 | 0 | 0 | 0 | 0 | 0 |
| 100 | 0.1 | 0.0 | 0.0 | 0 | 0 | 0 | 0.1 | 0.0 | 0.0 | 0 | 0 | 0 |
| 200 | 0.8 | 0.8 | 0.7 | 0 | 0 | 0 | 0.9 | 0.7 | 0.5 | 0 | 0 | 0 |
| 300 | 1.5 | 1.5 | 1.4 | 0.0 | 0 | 0 | 2.0 | 1.9 | 1.6 | 0.0 | 0.0 | 0.0 |
| 400 | 2.0 | 1.8 | 1.7 | 0.0 | 0.0 | 0.0 | 2.7 | 2.3 | 2.2 | 0.0 | 0.0 | 0.0 |
| 500 | 2.6 | 2.5 | 2.5 | 0.1 | 0.0 | 0.0 | 2.7 | 2.5 | 2.4 | 0.1 | 0.1 | 0.1 |
| | $\gamma_1=0.60$ $p_1=25\%$ $p_2=75\%$ | | | | | | $\gamma_1=0.80$ $p_1=25\%$ $p_2=75\%$ | | | | | |
| | $\alpha_{11}=2.44$ $\alpha_{01}=53$ $\alpha_{12}=1.03$ $\alpha_{02}=2.09$ | | | $\alpha_{11}=3.6$ $\alpha_{01}=94.6$ $\alpha_{12}=1.05$ $\alpha_{02}=2.96$ | | | $\alpha_{11}=2.44$ $\alpha_{01}=53$ $\alpha_{12}=1.03$ $\alpha_{02}=2.09$ | | | $\alpha_{11}=3.6$ $\alpha_{01}=94.6$ $\alpha_{12}=1.05$ $\alpha_{02}=2.96$ | | |
| *n* | *RM* | *WLS* | *WM* | *RM* | *WLS* | *WM* | *RM* | *WLS* | *WM* | *RM* | *WLS* | *WM* |
| 50 | 0 | 0 | 0 | 0 | 0 | 0 | 0 | 0 | 0 | 0 | 0 | 0 |
| 100 | 0.0 | 0.0 | 0.0 | 0 | 0 | 0 | 0.0 | 0.0 | 0.0 | 0 | 0 | 0 |
| 200 | 0.4 | 0.4 | 0.3 | 0 | 0 | 0 | 0.5 | 0.4 | 0.3 | 0 | 0 | 0 |
| 300 | 1.1 | 1.1 | 1.0 | 0.0 | 0.0 | 0.0 | 0.9 | 0.7 | 0.7 | 0.0 | 0.0 | 0.0 |
| 400 | 1.9 | 1.8 | 1.7 | 0.0 | 0.0 | 0.0 | 1.2 | 1.1 | 1.1 | 0.0 | 0.0 | 0.0 |
| 500 | 2.2 | 2.2 | 2.2 | 0.0 | 0.0 | 0.0 | 1.7 | 1.6 | 1.5 | 0.0 | 0.0 | 0.0 |

*RM*: regressions models. *WLS*: weighted least squares. *WM*: Wu method. The value 0.0 means $<0.05$.

Table 3. Powers (in %) of the global hypothesis test with a binary covariate.

| | $\tau_{11}=0.80$ $\upsilon_{11}=0.85$ $\tau_{21}=0.85$ $\upsilon_{21}=0.90$ $\tau_{12}=0.75$ $\upsilon_{12}=0.85$ $\tau_{22}=0.90$ $\upsilon_{22}=0.90$ $\tau_1\approx 0.80$ $\upsilon_1=0.85$ $\tau_2\approx 0.85$ $\upsilon_2=0.90$ | | | | | | | | | | | |
|---|---|---|---|---|---|---|---|---|---|---|---|---|
| | $\gamma_1=0.60$ $p_1=75\%$ $p_2=25\%$ | | | | | | $\gamma_1=0.80$ $p_1=75\%$ $p_2=25\%$ | | | | | |
| | $\alpha_{11}=1.00$ $\alpha_{01}=1.18$ $\alpha_{12}=1.24$ $\alpha_{02}=9.5$ | | | $\alpha_{11}=1.01$ $\alpha_{01}=1.32$ $\alpha_{12}=1.43$ $\alpha_{02}=16.3$ | | | $\alpha_{11}=1.00$ $\alpha_{01}=1.18$ $\alpha_{12}=1.24$ $\alpha_{02}=9.5$ | | | $\alpha_{11}=1.01$ $\alpha_{01}=1.32$ $\alpha_{12}=1.43$ $\alpha_{02}=16.3$ | | |
| *n* | *RM* | *WLS* | *WM* | *RM* | *WLS* | *WM* | *RM* | *WLS* | *WM* | *RM* | *WLS* | *WM* |

| $n$ | *RM* | *WLS* | *WM* | *RM* | *WLS* | *WM* | *RM* | *WLS* | *WM* | *RM* | *WLS* | *WM* |
|---|---|---|---|---|---|---|---|---|---|---|---|---|
| 50 | 0.0 | 0.0 | 0.0 | 0.0 | 0.0 | 0.0 | 0.1 | 0.0 | 0.0 | 0.0 | 0.0 | 0.0 |
| 100 | 5.3 | 4.3 | 3.4 | 3.8 | 2.7 | 2.1 | 7.0 | 4.8 | 4.3 | 5.3 | 3.0 | 2.7 |
| 200 | 40.3 | 39.1 | 37.4 | 53.8 | 51.9 | 50.0 | 44.5 | 40.1 | 38.7 | 60.8 | 54.9 | 53.8 |
| 300 | 71.2 | 70.5 | 69.7 | 90.4 | 89.6 | 89.1 | 74.3 | 70.5 | 69.8 | 92.1 | 89.8 | 89.4 |
| 400 | 87.3 | 87.0 | 86.5 | 98.6 | 98.6 | 98.5 | 88.8 | 86.5 | 86.2 | 99.0 | 98.5 | 98.4 |
| 500 | 94.7 | 94.6 | 94.4 | 99.9 | 99.9 | 99.9 | 95.7 | 94.9 | 94.8 | 99.9 | 99.8 | 99.8 |
| $\tau_{11}=0.80$ $\upsilon_{11}=0.80$ $\tau_{21}=0.85$ $\upsilon_{21}=0.95$; $\tau_{12}=0.80$ $\upsilon_{12}=0.85$ $\tau_{22}=0.90$ $\upsilon_{22}=0.95$; $\tau_1 \approx 0.80$ $\upsilon_1 \approx 0.81$ $\tau_2 \approx 0.90$ $\upsilon_2=0.95$ | | | | | | | | | | | | |
| | $\gamma_1=0.60$ $p_1=25\%$ $p_2=75\%$ | | | | | | $\gamma_1=0.80$ $p_1=25\%$ $p_2=75\%$ | | | | | |
| | $\alpha_{11}=1.09$ $\alpha_{01}=10.5$ $\alpha_{12}=1.00$ $\alpha_{02}=1.18$ | | | $\alpha_{11}=1.16$ $\alpha_{01}=18.1$ $\alpha_{12}=1.01$ $\alpha_{02}=1.32$ | | | $\alpha_{11}=1.09$ $\alpha_{01}=10.5$ $\alpha_{12}=1.00$ $\alpha_{02}=1.18$ | | | $\alpha_{11}=1.16$ $\alpha_{01}=18.1$ $\alpha_{12}=1.01$ $\alpha_{02}=1.32$ | | |
| $n$ | *RM* | *WLS* | *WM* | *RM* | *WLS* | *WM* | *RM* | *WLS* | *WM* | *RM* | *WLS* | *WM* |
| 50 | 1.7 | 1.1 | 0.7 | 1.6 | 1.0 | 0.4 | 5.7 | 3.0 | 2.2 | 5.6 | 2.8 | 2.1 |
| 100 | 53.2 | 48.8 | 44.2 | 57.4 | 51.9 | 47.4 | 75.6 | 67.4 | 65.1 | 80.6 | 71.1 | 69.3 |
| 200 | 98.2 | 98.0 | 97.7 | 99.5 | 99.4 | 99.3 | 99.8 | 99.7 | 99.7 | 100 | 100 | 100 |
| 300 | 100 | 100 | 100 | 100 | 100 | 100 | 100 | 100 | 100 | 100 | 100 | 100 |
| 400 | 100 | 100 | 100 | 100 | 100 | 100 | 100 | 100 | 100 | 100 | 100 | 100 |
| 500 | 100 | 100 | 100 | 100 | 100 | 100 | 100 | 100 | 100 | 100 | 100 | 100 |
| $\tau_{11}=0.80$ $\tau_{11}=0.90$ $\tau_{21}=0.95$ $\upsilon_{21}=0.80$; $\tau_{12}=0.80$ $\upsilon_{12}=0.95$ $\tau_{22}=0.95$ $\upsilon_{22}=0.90$; $\tau_1=0.80$ $\upsilon_1 \approx 0.94$ $\tau_2=0.95$ $\upsilon_2 \approx 0.85$ | | | | | | | | | | | | |
| | $\gamma_1=0.60$ $p_1=75\%$ $p_2=25\%$ | | | | | | $\gamma_1=0.80$ $p_1=75\%$ $p_2=25\%$ | | | | | |
| | $\alpha_{11}=1.00$ $\alpha_{01}=1.17$ $\alpha_{12}=1.09$ $\alpha_{02}=7.53$ | | | $\alpha_{11}=1.01$ $\alpha_{01}=1.31$ $\alpha_{12}=1.15$ $\alpha_{02}=12.75$ | | | $\alpha_{11}=1.00$ $\alpha_{01}=1.17$ $\alpha_{12}=1.09$ $\alpha_{02}=7.53$ | | | $\alpha_{11}=1.01$ $\alpha_{01}=1.31$ $\alpha_{12}=1.15$ $\alpha_{02}=12.75$ | | |
| $n$ | *RM* | *WLS* | *WM* | *RM* | *WLS* | *WM* | *RM* | *WLS* | *WM* | *RM* | *WLS* | *WM* |
| 50 | 3.0 | 1.8 | 2.2 | 2.9 | 1.4 | 1.8 | 6.9 | 3.8 | 5.3 | 6.0 | 3.0 | 4.3 |
| 100 | 80.2 | 78.1 | 73.6 | 85.2 | 82.9 | 78.3 | 90.2 | 87.7 | 85.9 | 93.7 | 91.1 | 90.1 |
| 200 | 99.9 | 99.9 | 99.9 | 100 | 99.9 | 99.9 | 100 | 99.9 | 99.9 | 100 | 100 | 100 |
| 300 | 100 | 100 | 100 | 100 | 100 | 100 | 100 | 100 | 100 | 100 | 100 | 100 |
| 400 | 100 | 100 | 100 | 100 | 100 | 100 | 100 | 100 | 100 | 100 | 100 | 100 |
| 500 | 100 | 100 | 100 | 100 | 100 | 100 | 100 | 100 | 100 | 100 | 100 | 100 |
| $\tau_{11}=0.85$ $\upsilon_{11}=0.80$ $\tau_{21}=0.95$ $\upsilon_{21}=0.90$; $\tau_{12}=0.85$ $\upsilon_{12}=0.75$ $\tau_{22}=0.80$ $\upsilon_{22}=0.90$; $\tau_1=0.85$ $\upsilon_1 \approx 0.78$ $\tau_2 \approx 0.81$ $\upsilon_2=0.90$ | | | | | | | | | | | | |
| | $\gamma_1=0.60$ $p_1=25\%$ $p_2=75\%$ | | | | | | $\gamma_1=0.80$ $p_1=25\%$ $p_2=75\%$ | | | | | |
| | $\alpha_{11}=1.25$ $\alpha_{01}=33$ $\alpha_{12}=1.00$ $\alpha_{02}=1.17$ | | | $\alpha_{11}=1.44$ $\alpha_{01}=58.6$ $\alpha_{12}=1.01$ $\alpha_{02}=1.31$ | | | $\alpha_{11}=1.25$ $\alpha_{01}=33$ $\alpha_{12}=1.00$ $\alpha_{02}=1.17$ | | | $\alpha_{11}=1.44$ $\alpha_{01}=58.6$ $\alpha_{12}=1.01$ $\alpha_{02}=1.31$ | | |
| $n$ | *RM* | *WLS* | *WM* | *RM* | *WLS* | *WM* | *RM* | *WLS* | *WM* | *RM* | *WLS* | *WM* |
| 50 | 0.4 | 0.2 | 0.2 | 0.3 | 0.1 | 0.1 | 1.0 | 0.3 | 0.4 | 0.8 | 0.2 | 0.2 |
| 100 | 34.0 | 28.0 | 25.9 | 35.9 | 26.9 | 26.0 | 44.5 | 33.5 | 33.4 | 48.9 | 34.9 | 35.5 |
| 200 | 93.1 | 91.6 | 90.9 | 97.9 | 96.8 | 96.4 | 96.0 | 93.8 | 93.4 | 99.0 | 97.9 | 97.9 |
| 300 | 99.7 | 99.6 | 99.5 | 100 | 100 | 100 | 99.8 | 99.6 | 99.6 | 100 | 100 | 100 |
| 400 | 100 | 100 | 100 | 100 | 100 | 100 | 100 | 100 | 100 | 100 | 100 | 100 |
| 500 | 100 | 100 | 100 | 100 | 100 | 100 | 100 | 100 | 100 | 100 | 100 | 100 |

*RM*: regressions models. *WLS*: weighted least squares. *WM*: Wu method.

Regarding the relative biases of the estimators obtained with the *RM* method and through *WLS*, Table 4 (A binary covariate) shows some results (rounded to four decimal places) for *PVs* of *Test* 1, and it is obtained that both estimators have practically the same relative biases (they vary from the sixth or seventh decimal place). Similar conclusions are obtained for the *PVs* of *Test* 2.

Table 4. Relative bias of estimators of *PVs*.

| | | | | | | | | |
|---|---|---|---|---|---|---|---|---|
| A binary covariate | | | | | | | | |
| $\tau_{11}=0.80$ $\upsilon_{11}=0.85$ $\tau_{21}=0.80$ $\upsilon_{21}=0.85$ $\tau_{12}=0.80$ $\upsilon_{12}=0.85$ $\tau_{22}=0.80$ $\upsilon_{22}=0.85$ | | | | | | | | |
| $\tau_1=\tau_2=0.80$ $\upsilon_1=\upsilon_2=0.85$ | | | | | | | | |
| | $\gamma_1=0.60$ $p_1=75\%$ $p_2=25\%$ | | | | $\gamma_1=0.80$ $p_1=75\%$ $p_2=25\%$ | | | |
| | $\alpha_{11}=1.00$ $\alpha_{01}=1.18$ $\alpha_{12}=1.52$ $\alpha_{02}=12.69$ | | | | $\alpha_{11}=1.00$ $\alpha_{01}=1.18$ $\alpha_{12}=1.52$ $\alpha_{02}=12.69$ | | | |
| | $\hat{\tau}_1$ | | $\hat{\upsilon}_1$ | | $\hat{\tau}_1$ | | $\hat{\upsilon}_1$ | |
| *n* | *RM* | *WLS* | *RM* | *WLS* | *RM* | *WLS* | *RM* | *WLS* |
| 50 | -0.1063 | -0.1063 | -0.1583 | -0.1583 | -0.0877 | -0.0877 | -0.2143 | -0.2143 |
| 100 | -0.0503 | -0.0503 | -0.0821 | -0.0821 | -0.0412 | -0.0412 | -0.1241 | -0.1241 |
| 200 | -0.0240 | -0.0240 | -0.0415 | -0.0415 | -0.0211 | -0.0211 | -0.0654 | -0.0654 |
| 300 | -0.0172 | -0.0172 | -0.0282 | -0.0282 | -0.0141 | -0.0141 | -0.0453 | -0.0453 |
| 400 | -0.0123 | -0.0123 | -0.0214 | -0.0214 | -0.0101 | -0.0101 | -0.0343 | -0.0343 |
| 500 | -0.0098 | -0.0098 | -0.0165 | -0.0165 | -0.0077 | -0.0077 | -0.0280 | -0.0280 |
| A covariate with three categories | | | | | | | | |
| $\tau_{1m}=0.85$ $\upsilon_{1m}=0.95$ | | | | | | | | |
| $\gamma_1=0.50$ $\gamma_2=0.30$ $\gamma_1=0.20$ $p_1=20\%$ $p_2=35\%$ $p_2=50\%$ | | | | | | | | |
| $\tau_1=\tau_2=0.85$ $\upsilon_1=\upsilon_2=0.95$ | | | | | | | | |
| | $\alpha_{11}=1.13$ $\alpha_{01}=14.72$ | | | | $\alpha_{11}=1.23$ $\alpha_{01}=25.7$ | | | |
| | $\alpha_{12}=1.05$ $\alpha_{02}=6.28$ | | | | $\alpha_{12}=1.09$ $\alpha_{02}=10.5$ | | | |
| | $\alpha_{13}=1.02$ $\alpha_{03}=3.46$ | | | | $\alpha_{13}=1.04$ $\alpha_{03}=5.43$ | | | |
| | $\hat{\tau}_1$ | | $\hat{\upsilon}_1$ | | $\hat{\tau}_1$ | | $\hat{\upsilon}_1$ | |
| *n* | *RM* | *WLS* | *RM* | *WLS* | *RM* | *WLS* | *RM* | *WLS* |
| 100 | -0.1368 | -0.0893 | -0.1368 | -0.0893 | -0.1368 | -0.0895 | -0.1368 | -0.0895 |
| 200 | -0.0740 | -0.0428 | -0.0740 | -0.0428 | -0.0732 | -0.0423 | -0.0732 | -0.0423 |
| 300 | -0.0489 | -0.0280 | -0.0489 | -0.0280 | -0.0494 | -0.0284 | -0.0494 | -0.0284 |
| 400 | -0.0375 | -0.0215 | -0.0375 | -0.0215 | -0.0381 | -0.0219 | -0.0381 | -0.0219 |
| 500 | -0.0305 | -0.0167 | -0.0305 | -0.0167 | -0.0302 | -0.0165 | -0.0302 | -0.0165 |
| A dummy binary covariate | | | | | | | | |
| | $\tau_1=\tau_2=0.90$ $\upsilon_1=\upsilon_2=0.80$ | | | | $\tau_1=\tau_2=0.75$ $\upsilon_1=\upsilon_2=0.85$ | | | |
| | $\gamma=0.50$ $p=25\%$ | | | | $\gamma=0.50$ $p=75\%$ | | | |
| | $\alpha_1=2.43$ $\alpha_0=53$ | | | | $\alpha_1=3.6$ $\alpha_0=94.6$ | | | |
| | $\hat{\tau}_1$ | | $\hat{\upsilon}_1$ | | $\hat{\tau}_1$ | | $\hat{\upsilon}_1$ | |
| *n* | *RM* | *WLS* | *RM* | *WLS* | *RM* | *WLS* | *RM* | *WLS* |
| 50 | -0.1032 | -0.1032 | -0.2020 | -0.2020 | -0.1999 | -0.1999 | -0.0894 | -0.0894 |
| 100 | -0.0477 | -0.0477 | -0.1182 | -0.1182 | -0.1245 | -0.1245 | -0.0418 | -0.0418 |
| 200 | -0.0232 | -0.0232 | -0.0649 | -0.0649 | -0.0707 | -0.0707 | -0.0212 | -0.0212 |
| 300 | -0.0155 | -0.0155 | -0.0431 | -0.0431 | -0.0473 | -0.0473 | -0.0132 | -0.0132 |
| 400 | -0.0116 | -0.0116 | -0.0344 | -0.0344 | -0.0363 | -0.0363 | -0.0104 | -0.0104 |
| 500 | -0.0095 | -0.0095 | -0.0271 | -0.0271 | -0.0304 | -0.0304 | -0.0077 | -0.0077 |

*RM*: regressions models. *WLS*: weighted least squares.

### *4.2. A covariate with three categories*

Simulation experiments were carried out to study the asymptotic behavior (type I error rates and the powers) of the global hypothesis test in the situation where a covariate with three categories is observed, applying the same methods as in the previous Section. In this case, three $2 \times 4$ tables are observed. These experiments have been designed in a similar way to those in Section 4.1, generating 10000 random samples of multinomial distributions whose probabilities have been calculated with the same equations (23) (considering $m = 1,2,3$). The sample size $n = \{100, 200, ..., 500\}$ has been considered. The sample size $n = 50$ has not been considered to avoid the presence of many frequencies equal to 0 (since three $2 \times 4$ tables are analyzed). For the *PVs* and covariances the same values have been considered as in Section 4.1. For $\boldsymbol{\gamma}$, two scenarios have been considered: $(\gamma_1 = 0.50, \gamma_2 = 0.30, \gamma_3 = 0.20)$ and $(\gamma_1 = 0.40, \gamma_2 = 0.35, \gamma_3 = 0.25)$. For the prevalence in each category of the covariate, two scenarios have been considered: $(p_1 = 20\%, p_2 = 35\%, p_3 = 50\%)$ and $(p_1 = 75\%, p_2 = 50\%, p_3 = 25\%)$. The experiments have also been designed in such a way that in all the generated samples the *PVs* and their variances-covariances can be estimated. Table 5 shows the results of the type I error rates obtained for different scenarios. As for a binary covariate, all three methods are conservative for the sample sizes considered and increasing covariance values implies a decrease in type I error rates. Table 6 shows the results of the powers obtained for different scenarios. The conclusions obtained are also similar to the case of a binary covariate. The *RM* method is more powerful than the *WLS* and *WM* methods. Depending on the scenario, with a moderate sample size, the *RM* method can have a power between 4% (approximately) and 10% greater than the *WM*. The increase in the covariances involves an increase in the power of each method, above all when the sample size is large.

Table 5. Type I error rates (in %) of the global hypothesis test with a covariate with three categories.

| $\tau_{1m} = 0.85 \;\; \upsilon_{1m} = 0.95$ |
| --- |
| $\gamma_1 = 0.50 \;\; \gamma_2 = 0.30 \;\; \gamma_1 = 0.20 \;\; p_1 = 20\% \;\; p_2 = 35\% \;\; p_2 = 50\%$ |
| $\tau_1 = \tau_2 = 0.85 \;\; \upsilon_1 = \upsilon_2 = 0.95$ |

| | $\alpha_{11}=1.13$ $\alpha_{01}=14.72$<br>$\alpha_{12}=1.05$ $\alpha_{02}=6.28$<br>$\alpha_{13}=1.02$ $\alpha_{03}=3.46$ | | | $\alpha_{11}=1.23$ $\alpha_{01}=25.7$<br>$\alpha_{12}=1.09$ $\alpha_{02}=10.5$<br>$\alpha_{13}=1.04$ $\alpha_{03}=5.43$ | | |
|---|---|---|---|---|---|---|
| *n* | *RM* | *WLS* | *WM* | *RM* | *WLS* | *WM* |
| 100 | 0 | 0 | 0 | 0 | 0 | 0 |
| 200 | 0.3 | 0.3 | 0.3 | 0 | 0 | 0 |
| 300 | 0.5 | 0.5 | 0.5 | 0 | 0 | 0 |
| 400 | 1.1 | 1.1 | 1.0 | 0 | 0 | 0 |
| 500 | 1.3 | 1.4 | 1.2 | 0.0 | 0.0 | 0.0 |

$\tau_{1m}=0.85$ $\upsilon_{1m}=0.95$

$\gamma_1=0.40$ $\gamma_2=0.35$ $\gamma_1=0.25$ $p_1=75\%$ $p_2=50\%$ $p_2=25\%$

$\tau_1=\tau_2=0.85$ $\upsilon_1=\upsilon_2=0.95$

| | $\alpha_{11}=1.00$ $\alpha_{01}=1.45$<br>$\alpha_{12}=1.02$ $\alpha_{02}=3.46$<br>$\alpha_{13}=1.09$ $\alpha_{03}=10.5$ | | | $\alpha_{11}=1.01$ $\alpha_{01}=1.81$<br>$\alpha_{12}=1.04$ $\alpha_{02}=5.43$<br>$\alpha_{13}=1.16$ $\alpha_{03}=18.1$ | | |
|---|---|---|---|---|---|---|
| *n* | *RM* | *WLS* | *WM* | *RM* | *WLS* | *WM* |
| 100 | 0 | 0 | 0 | 0 | 0 | 0 |
| 200 | 0.4 | 0.4 | 0.3 | 0 | 0 | 0 |
| 300 | 0.9 | 0.6 | 0.6 | 0 | 0 | 0 |
| 400 | 1.4 | 1.3 | 1.2 | 0 | 0 | 0 |
| 500 | 1.5 | 1.4 | 1.3 | 0.0 | 0.0 | 0.0 |

$\tau_{1m}=0.90$ $\upsilon_{1m}=0.85$

$\gamma_1=0.50$ $\gamma_2=0.30$ $\gamma_1=0.20$ $p_1=20\%$ $p_2=35\%$ $p_2=50\%$

$\tau_1=\tau_2=0.90$ $\upsilon_1=\upsilon_2=0.85$

| | $\alpha_{11}=2.17$ $\alpha_{01}=60.5$<br>$\alpha_{12}=1.23$ $\alpha_{02}=12.69$<br>$\alpha_{13}=1.10$ $\alpha_{03}=5.86$ | | | $\alpha_{11}=3.1$ $\alpha_{01}=108.1$<br>$\alpha_{12}=1.41$ $\alpha_{02}=22.04$<br>$\alpha_{13}=1.17$ $\alpha_{03}=9.74$ | | |
|---|---|---|---|---|---|---|
| *n* | *RM* | *WLS* | *WM* | *RM* | *WLS* | *WM* |
| 100 | 0 | 0 | 0 | 0 | 0 | 0 |
| 200 | 0.3 | 0.2 | 0.2 | 0 | 0 | 0 |
| 300 | 0.8 | 0.6 | 0.5 | 0 | 0 | 0 |
| 400 | 1.3 | 1.1 | 1.1 | 0.0 | 0.0 | 0.0 |
| 500 | 1.8 | 1.4 | 1.4 | 0.1 | 0.1 | 0.1 |

$\tau_{1m}=0.90$ $\upsilon_{1m}=0.85$

$\gamma_1=0.40$ $\gamma_2=0.35$ $\gamma_1=0.25$ $p_1=75\%$ $p_2=50\%$ $p_2=25\%$

$\tau_1=\tau_2=0.90$ $\upsilon_1=\upsilon_2=0.85$

| | $\alpha_{11}=1.02$ $\alpha_{01}=2.06$<br>$\alpha_{12}=1.10$ $\alpha_{02}=5.86$<br>$\alpha_{13}=1.54$ $\alpha_{03}=28.63$ | | | $\alpha_{11}=1.04$ $\alpha_{01}=2.91$<br>$\alpha_{12}=1.17$ $\alpha_{02}=9.74$<br>$\alpha_{13}=1.98$ $\alpha_{03}=50.73$ | | |
|---|---|---|---|---|---|---|
| *n* | *RM* | *WLS* | *WM* | *RM* | *WLS* | *WM* |
| 100 | 0 | 0 | 0 | 0 | 0 | 0 |
| 200 | 0.4 | 0.3 | 0.2 | 0 | 0 | 0 |
| 300 | 0.8 | 0.8 | 0.7 | 0 | 0 | 0 |
| 400 | 1.1 | 1.2 | 1.1 | 0.0 | 0.0 | 0.0 |
| 500 | 2.5 | 2.4 | 2.4 | 0.1 | 0.1 | 0.1 |

*RM*: regressions models. *WLS*: weighted least squares. *WM*: Wu method. The value 0.0 means $<0.05$.

Table 6. Powers (in %) of the global hypothesis test with a covariate with three categories.

$\tau_{11}=0.75$ $\upsilon_{11}=0.85$ $\tau_{21}=0.80$ $\upsilon_{21}=0.90$ $\tau_{12}=0.85$ $\upsilon_{12}=0.90$ $\tau_{22}=0.90$ $\upsilon_{22}=0.95$ $\tau_{13}=0.80$ $\upsilon_{13}=0.85$ $\tau_{23}=0.90$ $\upsilon_{23}=0.95$

$\gamma_1=0.50$ $\gamma_2=0.30$ $\gamma_3=0.20$ $p_1=20\%$ $p_2=35\%$ $p_3=50\%$

$\tau_1\approx 0.81$ $\tau_2\approx 0.89$ $\upsilon_1\approx 0.87$ $\upsilon_2\approx 0.94$

| | $\alpha_{11}=1.38$ $\alpha_{01}=14.5$<br>$\alpha_{12}=1.05$ $\alpha_{02}=7$<br>$\alpha_{13}=1.02$ $\alpha_{03}=2.82$ | | | $\alpha_{11}=1.68$ $\alpha_{01}=25.3$<br>$\alpha_{12}=1.09$ $\alpha_{02}=11.8$<br>$\alpha_{13}=1.04$ $\alpha_{03}=4.28$ | | |
|---|---|---|---|---|---|---|
| *n* | *RM* | *WLS* | *WM* | *RM* | *WLS* | *WM* |
| 100 | 2.4 | 1.3 | 1.0 | 1.8 | 0.9 | 0.7 |
| 200 | 33.4 | 30.0 | 28.7 | 45.4 | 36.9 | 34.3 |
| 300 | 69.4 | 66.2 | 65.0 | 87.0 | 83.3 | 82.4 |
| 400 | 87.7 | 85.2 | 84.5 | 97.1 | 96.2 | 96.0 |
| 500 | 95.1 | 93.5 | 93.4 | 99.9 | 99.8 | 99.8 |

| $\tau_{11}=0.80$ $\upsilon_{11}=0.80$ $\tau_{21}=0.80$ $\upsilon_{21}=0.90$ $\tau_{12}=0.85$ $\upsilon_{12}=0.90$ $\tau_{22}=0.90$ $\upsilon_{22}=0.95$ $\tau_{13}=0.80$ $\upsilon_{13}=0.85$ $\tau_{23}=0.90$ $\upsilon_{23}=0.95$ $\gamma_1=0.40$ $\gamma_2=0.35$ $\gamma_3=0.25$ $p_1=75\%$ $p_2=50\%$ $p_3=25\%$ $\tau_1\approx0.81$ $\tau_2\approx0.83$ $\upsilon_1\approx0.87$ $\upsilon_2\approx0.94$ | | | | | | |
|---|---|---|---|---|---|---|
| | $\alpha_{11}=1.00$ $\alpha_{01}=1.17$ $\alpha_{12}=1.02$ $\alpha_{02}=3.63$ $\alpha_{13}=1.09$ $\alpha_{03}=12.69$ | | | $\alpha_{11}=1.01$ $\alpha_{01}=1.31$ $\alpha_{12}=1.04$ $\alpha_{02}=5.73$ $\alpha_{13}=1.16$ $\alpha_{03}=22.04$ | | |
| *n* | *RM* | *WLS* | *WM* | *RM* | *WLS* | *WM* |
| 100 | 0.5 | 0.2 | 0.3 | 0.3 | 0.1 | 0.1 |
| 200 | 20.0 | 17.4 | 15.7 | 21.5 | 15.8 | 13.6 |
| 300 | 48.3 | 46.2 | 44.4 | 67.6 | 60.1 | 57.5 |
| 400 | 74.7 | 72.6 | 71.2 | 91.4 | 88.3 | 87.2 |
| 500 | 88.6 | 86.9 | 86.3 | 97.7 | 96.5 | 96.4 |
| $\tau_{11}=0.90$ $\upsilon_{11}=0.95$ $\tau_{21}=0.85$ $\upsilon_{21}=0.85$ $\tau_{12}=0.85$ $\upsilon_{12}=0.90$ $\tau_{22}=0.80$ $\upsilon_{22}=0.85$ $\tau_{13}=0.85$ $\upsilon_{13}=0.75$ $\tau_{23}=0.95$ $\upsilon_{23}=0.95$ $\gamma_1=0.50$ $\gamma_2=0.30$ $\gamma_3=0.20$ $p_1=20\%$ $p_2=35\%$ $p_3=50\%$ $\tau_1\approx0.85$ $\tau_2\approx0.90$ $\upsilon_1\approx0.85$ $\upsilon_2\approx0.89$ | | | | | | |
| | $\alpha_{11}=1.13$ $\alpha_{01}=23.17$ $\alpha_{12}=1.12$ $\alpha_{02}=5.78$ $\alpha_{13}=1.03$ $\alpha_{03}=4.5$ | | | $\alpha_{11}=1.23$ $\alpha_{01}=40.9$ $\alpha_{12}=1.21$ $\alpha_{02}=9.61$ $\alpha_{13}=1.05$ $\alpha_{03}=7.3$ | | |
| *n* | *RM* | *WLS* | *WM* | *RM* | *WLS* | *WM* |
| 100 | 2.1 | 0.9 | 0.9 | 1.3 | 0.6 | 0.6 |
| 200 | 14.6 | 11.1 | 10.1 | 17.5 | 11.0 | 10.6 |
| 300 | 31.6 | 25.7 | 25.2 | 39.5 | 29.9 | 29.1 |
| 400 | 47.7 | 42.3 | 41.2 | 58.8 | 48.5 | 48.1 |
| 500 | 60.6 | 54.2 | 53.2 | 71.6 | 64.7 | 64.2 |
| $\tau_{11}=0.85$ $\upsilon_{11}=0.95$ $\tau_{21}=0.80$ $\upsilon_{21}=0.85$ $\tau_{12}=0.85$ $\upsilon_{12}=0.90$ $\tau_{22}=0.75$ $\upsilon_{22}=0.85$ $\tau_{13}=0.80$ $\upsilon_{13}=0.90$ $\tau_{23}=0.90$ $\upsilon_{23}=0.90$ $\gamma_1=0.40$ $\gamma_2=0.35$ $\gamma_3=0.25$ $p_1=75\%$ $p_2=50\%$ $p_3=25\%$ $\tau_1\approx0.85$ $\tau_2\approx0.79$ $\upsilon_1\approx0.91$ $\upsilon_2\approx0.87$ | | | | | | |
| | $\alpha_{11}=1.00$ $\alpha_{01}=1.18$ $\alpha_{12}=1.05$ $\alpha_{02}=2.21$ $\alpha_{13}=1.22$ $\alpha_{03}=9.25$ | | | $\alpha_{11}=1.01$ $\alpha_{01}=1.32$ $\alpha_{12}=1.09$ $\alpha_{02}=3.19$ $\alpha_{13}=1.41$ $\alpha_{03}=15.85$ | | |
| *n* | *RM* | *WLS* | *WM* | *RM* | *WLS* | *WM* |
| 100 | 1.3 | 0.9 | 0.7 | 0.5 | 0.3 | 0.2 |
| 200 | 17.5 | 16.8 | 15.4 | 20.6 | 17.2 | 16.1 |
| 300 | 36.8 | 36.1 | 35.5 | 54.9 | 50.6 | 49.2 |
| 400 | 57.0 | 56.4 | 55.6 | 79.7 | 76.1 | 75.3 |
| 500 | 71.9 | 70.9 | 70.3 | 91.9 | 89.4 | 89.1 |

*RM*: regressions models. *WLS*: weighted least squares. *WM*: Wu method.

Regarding the estimators obtained using the *RM* and the *WLS*, Table 4 (A covariate with three categories) shows some results for the relative biases of the estimators of *PVs* of *Test* 1, showing that both types of estimators have practically the same relative bias. Similar conclusions are obtained for the *PVs* of *Test* 2.

The results of the simulation experiments have shown (as in the case of a binary covariate) that the *RM* method has a very similar type I error rate to the other two methods (*WLS* and *WM*), but has greater power. Therefore, when observing a covariate with three categories, the simultaneous comparison of the *PVs* of two *BDTs* should be performed using the *RM* method.

### *4.3. A binary dummy variable*

The simulation experiments carried out in Sections 4.1 and 4.2 have demonstrated that the method based on regression models (*RM*) has a greater power than those of the other two methods, and both have similar type I error rates. Furthermore, the *RM* and *WLS* methods require the observation of categorical covariates in all of the individuals in a sample. The question posed in this section is whether the *RM* and *WLS* methods can be applied when no covariate is observed. For this purpose, we propose the creation of a binary dummy variable and randomly assigning each individual to one of the two categories of this variable; next we can apply the *RM* or *WLS* method to compare the overall *PVs*. To study the asymptotic behavior of the *RM* and *WLS* methods in this situation, Monte Carlo simulation experiments were carried out, whose steps were the following:

1. We generated $N = 10,000$ random samples of multinomial distributions sized $n = \{50, 100, 200, ..., 500\}$ from a frequency table whose probabilities were calculated as

$$P(D=1, T_1 = j, T_2 = k) = p\left[\frac{\tau_1^j (\upsilon_1 - q)^j (\tau_1 + p)^{1-j} (1-\upsilon_1)^{1-j}}{p^j p^{1-j} Y_1^j Y_1^{1-j}} \times \right.$$

$$\left. \frac{\tau_2^k (\upsilon_2 - q)^k (\tau_2 + p)^{1-k} (1-\upsilon_2)^{1-k}}{p^k p^{1-k} Y_2^k Y_2^{1-k}} + \delta_{ik}\varepsilon_1 \right],$$

$$P(D=0, T_1 = j, T_2 = k) = q\left[\frac{(1-\tau_1)^j (\upsilon_1 - q)^j (\tau_1 - p)^{1-j} \upsilon_1^{1-j}}{q^j q^{1-j} Y_{1m}^j Y_{1m}^{1-j}} \times \right.$$

$$\left. \frac{(1-\tau_2)^k (\upsilon_2 - q)^k (\tau_2 - p)^{1-k} \upsilon_2^{1-k}}{q^k q^{1-k} Y_2^k Y_2^{1-k}} + \delta_{jk}\varepsilon_0 \right],$$

with

$$\alpha_1 = \frac{f}{\max\left\{\frac{\tau_1(\upsilon_1 - q)}{pY}, \frac{\tau_2(\upsilon_2 - q)}{pY_2}\right\}} + 1 - f$$

and

$$\alpha_0 = \frac{f}{\max\left\{\left(1 - \frac{(\tau_1 - p)\upsilon_1}{qY_1}\right), \left(1 - \frac{(\tau_2 - p)\upsilon_2}{qY_2}\right)\right\}} + 1 - f .$$

As values of the prevalence $p$ we took 25% and 75%, and as *PVs* we took $\{0.75, 0.80, ..., 0.95\}$. As values of the covariances $\alpha_1$ and $\alpha_0$ we took intermediate and high values, in a similar way to in Section 4.

2. Once each sample sized $n$ is generated, each individual has been randomly assigned to one of the categories of the dummy variable, generating for this purpose a Bernoulli random variable with probability $\gamma$, taking $\gamma = \{0.5, 0.7, 0.9\}$.

3. Once the $N$ samples in each scenario are generated, we calculated the type I error rates (or powers, depending on the type of scenario) of the global hypothesis test applying *RM*, *WLS* and *WM*. We also calculated the relative biases of the estimators of the overall *PVs* obtained through *RM* and *WLS*.

Table 7 shows the rates of the type I error rates of the *RM*, *WLS* and *WM* for different scenarios, intermediate values of the covariances $\varepsilon_1$ and $\varepsilon_0$, and for $\gamma = \{0.5, 0.7, 0.9\}$. The three methods have a very similar type I error rate, and these three methods are conservative. The probability $\gamma$ of the Bernoulli variable with which each individual is assigned a pattern of the covariate has practically no effect on the type I error rate of each method, and in all cases they are very similar. Similar conclusions are obtained when the covariances $\varepsilon_1$ and $\varepsilon_0$ are large.

Table 8 shows the powers of the three methods for different scenarios and also for intermediate values of the covariances and $\gamma = \{0.5, 0.7, 0.9\}$. In general terms, the power of the *RM* method is always higher than the powers of the *WLS* and *WM* methods, above all when the sample size is moderate ($n = 100 - 200$) or large $(n \geq 300)$. In some scenarios the power of *RM* may be more than 10% higher than the power of *WM* when the sample size is moderate and even more than 5% with a large sample size. The probability $\gamma$ has practically no effect on the power of each method. Similar conclusions are obtained when the covariances between the two *BDTs* are high.

Regarding the relative biases of the estimators obtained with *RM* and through *MLE* (see Appendix I), Table 4 (Binary dummy covariate) shows some results for *PVs* of *Test* 1, and it is obtained that both estimators have practically the same relative biases (they vary from the sixth or seventh decimal place). Similar conclusions are obtained for the *PVs* of *Test* 2.

Table 7. Type I error rates (in %) of the global hypothesis test with a dummy binary covariate.

| $\tau_1 = \tau_2 = 0.90$ $\upsilon_1 = \upsilon_2 = 0.80$ $p = 25\%$ <br> $\alpha_1 = 0.097$ $\alpha_0 = 0.005$ | | | | | | | | | |
|---|---|---|---|---|---|---|---|---|---|
| | $\gamma = 0.50$ | | | $\gamma = 0.70$ | | | $\gamma = 0.90$ | | |
| *n* | *RM* | *WLS* | *WM* | *RM* | *WLS* | *WM* | *RM* | *WLS* | *WM* |
| 50 | 0 | 0 | 0 | 0 | 0 | 0 | 0 | 0 | 0 |
| 100 | 0.0 | 0 | 0 | 0.0 | 0 | 0 | 0.0 | 0 | 0 |
| 200 | 0.4 | 0.2 | 0.2 | 0.5 | 0.2 | 0.2 | 0.5 | 0.2 | 0.2 |
| 300 | 1.0 | 0.5 | 0.6 | 1.0 | 0.6 | 0.5 | 0.9 | 0.5 | 0.5 |
| 400 | 1.2 | 0.6 | 0.6 | 1.5 | 0.9 | 0.9 | 1.4 | 0.8 | 0.8 |
| 500 | 1.7 | 1.1 | 1.1 | 1.6 | 1.0 | 1.0 | 1.6 | 1.0 | 1.0 |
| $\tau_1 = \tau_2 = 0.90$ $\upsilon_1 = \upsilon_2 = 0.80$ $p = 75\%$ <br> $\alpha_1 = 0.027$ $\alpha_0 = 0.108$ | | | | | | | | | |
| | $\gamma = 0.50$ | | | $\gamma = 0.70$ | | | $\gamma = 0.90$ | | |
| *n* | *RM* | *WLS* | *WM* | *RM* | *WLS* | *WM* | *RM* | *WLS* | *WM* |
| 50 | 0 | 0 | 0 | 0 | 0 | 0 | 0.0 | 0 | 0 |
| 100 | 0.2 | 0.1 | 0.1 | 0.2 | 0.1 | 0.1 | 0.2 | 0.1 | 0.0 |
| 200 | 1.2 | 1.0 | 0.9 | 1.3 | 1.0 | 0.8 | 1.3 | 1.0 | 0.8 |
| 300 | 2.0 | 1.9 | 1.7 | 2.3 | 2.0 | 1.8 | 2.5 | 2.2 | 2.0 |
| 400 | 2.7 | 2.5 | 2.2 | 2.7 | 2.3 | 2.1 | 2.9 | 2.7 | 2.4 |
| 500 | 3.2 | 2.6 | 2.4 | 3.2 | 3.0 | 2.8 | 3.3 | 3.1 | 2.9 |

$\tau_1 = \tau_2 = 0.85$ $\upsilon_1 = \upsilon_2 = 0.90$ $p = 25\%$

$\alpha_1 = 0.19$ $\alpha_0 = 0.019$

| | $\gamma = 0.50$ | | | $\gamma = 0.70$ | | | $\gamma = 0.90$ | | |
|---|---|---|---|---|---|---|---|---|---|
| *n* | *RM* | *WLS* | *WM* | *RM* | *WLS* | *WM* | *RM* | *WLS* | *WM* |
| 50 | 0 | 0 | 0 | 0 | 0 | 0 | 0 | 0 | 0 |
| 100 | 0.2 | 0.0 | 0.0 | 0.1 | 0.0 | 0.0 | 0.1 | 0 | 0 |
| 200 | 0.9 | 0.6 | 0.5 | 1.1 | 0.7 | 0.6 | 0.8 | 0.5 | 0.4 |
| 300 | 2.0 | 1.5 | 1.3 | 1.8 | 1.3 | 1.2 | 1.7 | 1.4 | 1.2 |
| 400 | 2.4 | 2.1 | 1.8 | 2.7 | 2.1 | 1.9 | 2.4 | 1.9 | 1.7 |
| 500 | 3.0 | 2.6 | 2.4 | 2.9 | 2.4 | 2.2 | 2.7 | 2.3 | 2.2 |

$\tau_1 = \tau_2 = 0.75$ $\upsilon_1 = \upsilon_2 = 0.85$ $p = 75\%$

$\alpha_1 = 0.009$ $\alpha_0 = 0.125$

| | $\gamma = 0.50$ | | | $\gamma = 0.70$ | | | $\gamma = 0.90$ | | |
|---|---|---|---|---|---|---|---|---|---|
| *n* | *RM* | *WLS* | *WM* | *RM* | *WLS* | *WM* | *RM* | *WLS* | *WM* |
| 50 | 0 | 0 | 0 | 0 | 0 | 0 | 0 | 0 | 0 |
| 100 | 0.1 | 0.0 | 0.0 | 0.2 | 0.0 | 0.0 | 0.1 | 0.0 | 0.0 |
| 200 | 0.9 | 0.4 | 0.4 | 0.8 | 0.4 | 0.3 | 0.9 | 0.4 | 0.5 |
| 300 | 1.5 | 1.0 | 0.9 | 1.4 | 0.8 | 0.8 | 1.4 | 1.0 | 1.0 |
| 400 | 1.9 | 1.4 | 1.4 | 2.0 | 1.4 | 1.3 | 2.0 | 1.5 | 1.4 |
| 500 | 2.0 | 1.6 | 1.5 | 2.1 | 1.7 | 1.6 | 2.1 | 1.6 | 1.6 |

*RM*: regressions models. *WLS*: weighted least squares. *WM*: Wu method. The value 0.0 means $< 0.05$.

The results of the simulation experiments carried out demonstrated that the *RM* method can be applied when no covariate is observed in the individuals in the sample. For this purpose, a binary dummy variable is created and each individual in the sample is randomly assigned a pattern of this variable. This method, which has a type I error rate similar to that of *WM*, has a higher power than *WM*, above all when the sample size is moderate or large.

Table 8. Power (in %) of the global hypothesis test with a dummy binary covariate.

$\tau_1 = 0.85$ $\upsilon_1 = 0.80$ $\tau_2 = 0.80$ $\upsilon_2 = 0.85$ $p = 25\%$

$\alpha_1 = 0.066$ $\alpha_0 = 0.007$

| | $\gamma = 0.50$ | | | $\gamma = 0.70$ | | | $\gamma = 0.90$ | | |
|---|---|---|---|---|---|---|---|---|---|
| *n* | *RM* | *WLS* | *WM* | *RM* | *WLS* | *WM* | *RM* | *WLS* | *WM* |
| 50 | 0.2 | 0.0 | 0.1 | 0.2 | 0.1 | 0.1 | 0.2 | 0.1 | 0.1 |
| 100 | 11.0 | 6.9 | 10.0 | 10.8 | 6.9 | 10.0 | 10.8 | 7.0 | 10.0 |
| 200 | 64.6 | 60.1 | 62.8 | 64.7 | 60.0 | 62.9 | 64.7 | 60.1 | 62.7 |
| 300 | 91.1 | 89.8 | 90.1 | 91.1 | 89.7 | 90.1 | 91.3 | 90.1 | 90.5 |
| 400 | 98.2 | 97.8 | 97.9 | 97.9 | 97.6 | 97.6 | 97.9 | 97.8 | 97.9 |
| 500 | 99.6 | 99.5 | 99.5 | 99.6 | 99.6 | 99.6 | 99.6 | 99.6 | 99.6 |

$\tau_1 = 0.85 \;\; \upsilon_1 = 0.80 \;\; \tau_2 = 0.80 \;\; \upsilon_2 = 0.85 \;\; p = 75\%$

$\alpha_1 = 0.007 \;\; \alpha_0 = 0.066$

| | $\gamma = 0.50$ | | | $\gamma = 0.70$ | | | $\gamma = 0.90$ | | |
|---|---|---|---|---|---|---|---|---|---|
| *n* | *RM* | *ML* | *WM* | *RM* | *ML* | *WM* | *RM* | *ML* | *WM* |
| 50 | 0.2 | 0.1 | 0.1 | 0.2 | 0.0 | 0.1 | 0.2 | 0.1 | 0.1 |
| 100 | 11.5 | 7.3 | 10.5 | 11.3 | 7.1 | 10.3 | 10.6 | 6.9 | 9.9 |
| 200 | 65.0 | 60.0 | 62.9 | 64.5 | 59.3 | 62.2 | 65.3 | 60.8 | 63.5 |
| 300 | 91.5 | 90.1 | 90.6 | 91.0 | 89.8 | 90.2 | 90.9 | 89.7 | 90.1 |
| 400 | 97.9 | 97.6 | 97.7 | 97.9 | 97.7 | 97.8 | 98.0 | 97.9 | 98.0 |
| 500 | 99.7 | 99.6 | 99.6 | 99.6 | 99.6 | 99.6 | 99.7 | 99.6 | 99.6 |

$\tau_1 = 0.80 \;\; \upsilon_1 = 0.90 \;\; \tau_2 = 0.85 \;\; \upsilon_2 = 0.95 \;\; p = 25\%$

$\alpha_1 = 0.051 \;\; \alpha_0 = 0.024$

| | $\gamma = 0.50$ | | | $\gamma = 0.70$ | | | $\gamma = 0.90$ | | |
|---|---|---|---|---|---|---|---|---|---|
| *n* | *RM* | *WLS* | *WM* | *RM* | *WLS* | *WM* | *RM* | *WLS* | *WM* |
| 50 | 0.1 | 0.0 | 0.0 | 0.0 | 0.0 | 0.0 | 0.1 | 0.0 | 0.0 |
| 100 | 6.0 | 3.1 | 2.4 | 5.6 | 2.9 | 2.3 | 6.1 | 3.2 | 2.5 |
| 200 | 40.2 | 31.2 | 29.5 | 41.1 | 31.6 | 29.9 | 39.6 | 30.5 | 28.9 |
| 300 | 68.9 | 60.1 | 59.1 | 69.4 | 60.0 | 58.8 | 68.4 | 59.8 | 58.9 |
| 400 | 85.2 | 78.8 | 78.3 | 86.0 | 79.5 | 78.8 | 85.5 | 79.4 | 78.9 |
| 500 | 93.4 | 89.6 | 89.3 | 93.8 | 89.7 | 89.3 | 93.4 | 89.7 | 89.3 |

$\tau_1 = 0.80 \;\; \upsilon_1 = 0.90 \;\; \tau_2 = 0.85 \;\; \upsilon_2 = 0.95 \;\; p = 75\%$

$\alpha_1 = 0.004 \;\; \alpha_0 = 0.068$

| | $\gamma = 0.50$ | | | $\gamma = 0.70$ | | | $\gamma = 0.90$ | | |
|---|---|---|---|---|---|---|---|---|---|
| *n* | *RM* | *WLS* | *WM* | *RM* | *WLS* | *WM* | *RM* | *WLS* | *WM* |
| 50 | 0.1 | 0.0 | 0.0 | 0.1 | 0.0 | 0.0 | 0.1 | 0.0 | 0.0 |
| 100 | 10.3 | 5.5 | 5.5 | 9.9 | 5.4 | 5.3 | 9.5 | 5.2 | 5.1 |
| 200 | 52.0 | 42.4 | 41.8 | 52.7 | 42.7 | 42.0 | 51.5 | 41.8 | 41.2 |
| 300 | 78.5 | 72.8 | 72.6 | 79.3 | 73.0 | 72.4 | 79.1 | 72.9 | 72.6 |
| 400 | 91.9 | 89.3 | 88.9 | 92.3 | 89.6 | 89.3 | 92.3 | 89.6 | 89.3 |
| 500 | 97.0 | 95.9 | 95.8 | 97.2 | 96.2 | 96.2 | 97.4 | 96.2 | 96.2 |

*RM*: regressions models. *WLS*: weighted least squares. *WM*: Wu method. The value 0.0 means $< 0.05$.

## 5. Functions in *R*

Two functions in *R* [16] were written to solve the problems studied. The first function called *RMCPV* (Regression Models to Compare Predictive Values), solves the global hypothesis test when a binary covariate is observed in all individuals; the second function, called *RMCPVDC* (Regression Models to Compare Predictive Values with a Dummy Covariate), solves the global test when a binary dummy covariate is created. The first function is executed with the sentence

$$RMCPV\left(s_{111}, s_{101}, s_{011}, s_{001}, r_{111}, r_{101}, r_{011}, r_{001}, s_{112}, s_{102}, s_{012}, s_{002}, r_{112}, r_{102}, r_{011}, r_{002}\right)$$

and the second one with the sentence

$$RMCPVDC\left(s_{11}, s_{10}, s_{01}, s_{00}, r_{11}, r_{10}, r_{01}, r_{00}\right).$$

Both functions check that the values of the frequencies are valid ones (there are no negative values, decimals,…) and solve the global test with $\alpha = 5\%$. The function *RMCPVDC* uses the value $\gamma = 0.5$ to generate the random value of Bernoulli variable. The two programmes provide the estimations of the overall *PVs* and their standard errors, the estimated variance-covariance matrix and the result of the global test. If the global test is significant with $\alpha = 5\%$, the two individual hypothesis tests are solved and the Holm method [20] is applied to investigate the causes of the significance. The two functions are available as supplementary material of the article.

## 6. Applications

The results obtained were applied to two real examples: one linked to the diagnosis of colorectal cancer and the other to the diagnosis of malaria.

### *6.1. Diagnosis of colorectal cancer*

Colorectal cancer (*CRC*) is the third most common type of cancer in the world and represents a major public health problem in the Western world. Different Studies have shown that diabetes is a factor that increases the risk of suffering from *CRC*. The diagnosis of *CRC* can be made through laboratory tests such as Fecal Immunochemical Testing (*FIT*) and Fecal Occult Blood Testing (*FOBT*), and biopsies are used as a gold standard. Table 9 (Study of *CRC*) shows the data for 168 men depending on whether they were diabetic or not. Executing the function *RMCPV* with the sentence

$$RMCPV(51,13,1,9,3,1,1,34,17,5,0,4,1,0,1,27)$$

it is obtained that $\hat{\tau}_1 = 94.5\%$, $\hat{\upsilon}_1 = 81.8\%$, $\hat{\tau}_2 = 92.0\%$ and $\hat{\upsilon}_2 = 66.7\%$. The test statistic for the global test is $Q = 18.1077$ and the $p-value = 1.2\times 10^{-4}$. Therefore, setting $\alpha = 5\%$ the joint equality of the two positive *PVs* and of the two negative *PVs* is rejected. As the global test is significant, the causes of the significance are studied solving the individual hypothesis tests and adjusting the *p-values* applying the Holm method [20]. For the test $H_0 : \tau_1 = \tau_2$ the test statistic is $z = 1.0977$ and $two-sided\ p-value = 0.2723$, and for the test $H_0 : \upsilon_1 = \upsilon_2$ the test statistic is $z = 4.255$ and $two-sided\ p-value = 0$. Applying the Holm method [20] the same *p-values* as before are obtained. Setting $\alpha = 5\%$ the equality of the positive *PVs* of the *FIT* and of the *FOBT* is not rejected and the negative *PV* of the *FIT* is significantly greater than that of the *FOBT*. Therefore, individuals in the population in whom the *FIT* gives a positive result have the same probability of having *CRC* as individuals in the same population in whom the *FOBT* gives a positive result. However, individuals in the population in whom the *FIT* gives a negative result have a higher probability of not having *CRC* than individuals in the same population in whom the *FOBT* gives a negative result.

If in this example the global test is solved without considering diabetes, with the method of Wu method [15] (see Appendix II) we obtain that $\chi^2 = 15.5439$ and $p-value = 4.2\times 10^{-4}$. Therefore, applying this method the same conclusion is obtained, although as the value of the test statistic is smaller than that obtained with the regression method, the power of the global hypothesis test is greater with the regression method than with the method of Wu [15]. We will study the causes of the significance by applying the Kosinski method [8] and adjusting the *p*-values with the Holm method [20]. For the comparison test of the two positive *PVs*, the test statistic is 1.587 and the adjusted two-tailed *p*-value is 0.208, therefore the equality of the two positive *PVs* is not rejected. For the comparison test of the two negative *PVs*, the test statistic is 15.682 and the adjusted two-tailed *p*-value is 0, therefore the equality of the two negative *PVs* is rejected.

Table 9. Data from *CRC* and malaria studies.

| | Study of *CRC* | | | |
|---|---|---|---|---|
| | Diabetics | | | |
| | *Positive Biopsy* | | *Negative Biopsy* | |
| | *FOBT* | | *FOBT* | |
| *FIT* | Positive | Negative | Positive | Negative |
| Positive | 51 | 13 | 3 | 1 |
| Negative | 1 | 9 | 1 | 34 |
| | Non-diabetics | | | |
| | *Positive Biopsy* | | *Negative Biopsy* | |
| | *FOBT* | | *FOBT* | |
| *FIT* | Positive | Negative | Positive | Negative |
| Positive | 17 | 5 | 1 | 0 |
| Negative | 0 | 4 | 1 | 27 |
| | Study of malaria | | | |
| | *Positive PCR* | | *Negative PCR* | |
| | *HRP2* | | *HRP2* | |
| *EMT* | Positive | Negative | Positive | Negative |
| Positive | 41 | 0 | 5 | 1 |
| Negative | 40 | 8 | 24 | 181 |

*6.2. Diagnosis of malaria*

The results obtained were applied to the study of Batwala et al [24] on the diagnosis of malaria. Batwala et al [24] applied the Expert Microscopy Test (*EMT*) and the *HRP2*-Based Rapid Diagnostic Test (*HRP2*) to a sample of 300 individuals using as the *GS* a *PCR*. The data from the study is shown in Table 9 (Study of Malaria). Executing the function *RMCPVDC* with the sentence

$$RMCPVDC\left(41,0,40,8,5,1,24,181\right)$$

it is obtained that $\hat{\tau}_1 = 87.2\%$, $\hat{\upsilon}_1 = 81.0\%$, $\hat{\tau}_2 = 73.6\%$ and $\hat{\upsilon}_2 = 95.8\%$. The test statistic for the global test is $Q = 78.3178$ and the $p-value = 0$. Therefore, setting $\alpha = 5\%$ the joint equality of the two positive *PVs* and of the two negative *PVs* is rejected. As the global test is significant, the causes of the significance are studied solving the individual hypothesis tests and adjusting the *p-values* applying the Holm method [20]. For the test $H_0 : \tau_1 = \tau_2$ the test statistic is $z = 2.8132$ and $two-sided\ p-value = 0.0049$, and for the test $H_0 : \upsilon_1 = \upsilon_2$ the test statistic is $z = -6.5728$ and $two-sided\ p-value = 0$. Applying the Holm method [20] the same *p-values* as before are obtained. Therefore, the positive *PV* of the *EMT* is significantly greater than that of the *HRP2* and the negative *PV* of the *HRP2* is significantly greater than that of the *EMT*. Therefore, individuals in the population in whom the *EMT* gives a positive result have a higher probability of having malaria as individuals in the same population in whom the *HRP2* gives a positive result. However, individuals in the population in whom the *HRP2* gives a negative result have a higher probability of not having *malaria* than individuals in the same population in whom the *EMT* gives a negative result.

Solving the global test applying the method of Wu [15], we obtain that $\chi^2 = 61.16$ and $p-value = 5.24 \times 10^{-14}$. Therefore, applying this method the same conclusion is obtained, and with respect to the power of the global hypothesis test, the same conclusion is also obtained. We will study the causes of the significance by applying the Kosinski method [8] and adjusting the *p*-values with the Holm method [20]. For the comparison test of the two positive *PVs*, the test statistic is 5.595 and the adjusted two-tailed *p*-value is 0, therefore the equality of the two positive *PVs* is rejected. For the comparison test of the two negative *PVs*, the test statistic is 33.871 and the adjusted two-tailed *p*-value is 0, therefore the equality of the two negative *PVs* is rejected.

## 7. Discussion

Two methods have been proposed to study a global hypothesis test of simultaneous comparison of the *PVs* of two *BDTs* when in all of the individuals in the sample categorical covariates are observed. The first method solves the global test applying logistic regression models and logit multinomial models (*RM*), and the second model applies weighted least squares (*WLS*) method for analysis of categorical data. In both models the test statistic is distributed according to a chi-square distribution with two degrees of freedom when the sample size is large. When the global hypothesis test is significant at the fixed $\alpha$ , the causes of significance are investigated by solving the individual hypothesis tests and applying the Holm [20] multiple comparison method, which is less conservative than the classic method of Bonferroni.

In both proposed methods, *RM* and *WLS*, the parameter $\boldsymbol{\gamma}$ is estimated in the same way. However, this is not the case with the parameters $\boldsymbol{\phi}$ and $\boldsymbol{\varphi}$. In the *WLS* model, these two parameters are estimated as multinomial proportions, each based on a multinomial distribution (both being independent). In the *RM* model, the parameter $\boldsymbol{\phi}$ is estimated using a logistic regression model and the parameter $\boldsymbol{\varphi}$ using a multinomial logit regression model. While the *WLS* model can only be used when the covariates are categorical, both regression models (logistic regression and logit multinomial regression) can also be used when the covariates are continuous. We have compared the estimators of the *PVs* obtained with both models (*RM* and *WLS*) through simulation experiments, calculating the relative biases of both types of estimators, obtaining that both methods give rise to estimates that are practically equal. Thus, if the covariates are continuous, they can be easily included in the logistic regression model and in the multinomial logit model. If the joint distribution of the covariates is known then $\gamma_m = P(\mathbf{X} = \mathbf{x}_m)$ can be calculated from the joint distribution function. For example, if a continuous covariate

(with a certain probability distribution, for example a normal distribution) is observed, then

$$\gamma_m = P(x_{1m} \leq X \leq x_{2m}) = \int_{x_{1m}}^{x_{2m}} f(x)\,dx,$$

where $f(x)$ is the probability density function of the normal distribution. The variance of $\gamma_m$ can be estimated by applying the bootstrap method. In the case of two or more continuous covariates, the solution to the problem is analogous considering the joint probability density function. Regarding the intervals for calculating probabilities of the covariates, the clinical normal ranges (or reference ranges) of the covariates can be used. For example, if the covariate is total cholesterol, the reference ranges are: less than 200 (healthy), between 200 and 239 (high risk), and 240 and above (dangerous).

Simulation experiments have been carried out to study the asymptotic behavior of both methods (*RM* and *WLS*) in two different situations, when a binary covariate is observed and when a covariate with three categories is observed, and both methods have been compared with the Wu method [15] (which ignores the covariates). The experiments have shown that the method based on regression models has a higher power than the method based on weighted least squares, with both methods having very similar type I error rates. The method based on regression models and the Wu method [15] are conservative methods (for the sample sizes considered) when a binary covariate (or a covariate with three categories) is observed. In this situation, the method based on regression models is more powerful than method of Wu [15] when the sample size is moderate (or even large, depending on the type of covariate). Therefore, it is recommended to apply the method based on regression models to simultaneously compare the *PVs* of two *BDTs*.

The regression models cannot be applied when there are many frequencies equal to 0. A necessary condition to be able to apply the regression model is that $s_{11m} + r_{11m} \geq 1$ and $s_{00m} + r_{00m} \geq 1$ (or $s_{10m} + r_{10m} \geq 1$ and $s_{01m} + r_{01m} \geq 1$) for $m = 1,...,M$. Therefore it cannot happen that in all covariate patterns it is verified that $s_{11m} + r_{11m} = s_{00m} + r_{00m} = 0$ (or $s_{10m} + r_{10m} = s_{01m} + r_{01m} = 0$). If the above is verified, then the logistic regression model and the multinomial logit model cannot be applied. Furthermore, it is obvious that there are no individuals without the disease or no individuals with the disease.

Furthermore, other simulation experiments were carried out when no covariate is observed in the individuals in the sample. For this purpose, a dummy binary variable was created, and the two methods proposed (regression models and weighted least squares) and each individual in the sample was randomly assigned a pattern of the covariate. The experiments showed that the method based on the regression models has a greater power than the other two methods (weighted least squares and the Wu method [15]), and the type I error rates are very similar. Therefore, when no binary covariate is observed, it is recommendable to apply the method based on regression models creating a dummy binary variable.

The method based on regression models can therefore be applied in two different situations: when categorical covariates are observed in all individuals in the sample and when no covariates are observed. In both situations, the same parametric model is used.

Two functions in *R* [16] were written that allow us to solve the problems studied and they are available as supplementary material for the article.

In the proposed regression model, parameters are modeled using logistic and multinomial logit regressions. An alternative to these models is to apply *GEE* models [6, 8]. Therefore, future work can address the extension of the results of Leisenring et al. [6] and Kosinski [8] to simultaneously compare *PVs* in the presence of categorical covariates.

Takahashi et al [25] have studied the comparison of the *PVs* of two *BDTs* by combining a superiority test and a non-inferiority test. Future work will focus on extending the results of Takahashi et al to the case where categorical covariates are observed in all individuals in the sample.

## Appendix I

When the covariates are not considered, the *MLEs* of the *PVs* (calculated from a single $2\times 4$ table) for *Test* 1 are

$$\hat{\tau}_1 = \frac{s_{10}+s_{11}}{n_{10}+n_{11}} \text{ and } \hat{\upsilon}_1 = \frac{r_{00}+r_{01}}{n_{00}+n_{01}}, \tag{24}$$

and for *Test* 2

$$\hat{\tau}_2 = \frac{s_{01}+s_{11}}{n_{01}+n_{11}} \text{ and } \hat{\upsilon}_2 = \frac{r_{00}+r_{10}}{n_{00}+n_{10}}, \tag{25}$$

where $n_{jk} = s_{jk} + r_{jk}$. Applying the delta method, the estimated variances-covariances are [15]

$$\hat{V}ar(\hat{\tau}_1) = \frac{(s_{10}+s_{11})(r_{10}+r_{11})}{(n_{10}+n_{11})^3},\ \hat{V}ar(\hat{\upsilon}_1) = \frac{(s_{00}+s_{01})(r_{00}+r_{01})}{(n_{00}+n_{01})^3},$$

$$\hat{V}ar(\hat{\tau}_2) = \frac{(s_{01}+s_{11})(r_{01}+r_{11})}{(n_{01}+n_{11})^3},\ \hat{V}ar(\hat{\upsilon}_2) = \frac{(s_{00}+s_{10})(r_{00}+r_{10})}{(n_{00}+n_{10})^3},$$

$$\hat{C}ov(\hat{\tau}_1,\hat{\tau}_2) = \frac{s_{11}r_{01}r_{10} + r_{11}\left[s_{01}(s_{10}+s_{11}) + s_{11}(r_{01}+r_{10}+r_{11}+s_{10}+s_{11})\right]}{(n_{01}+n_{11})^2(n_{10}+n_{11})^2},$$

$$\hat{C}ov(\hat{\tau}_1,\hat{\upsilon}_2) = -\frac{s_{10}r_{00}(r_{10}+r_{11}) + r_{10}\left[s_{11}(s_{00}+s_{10}) + s_{10}(s_{00}+s_{10}+r_{10}+r_{11})\right]}{(n_{00}+n_{10})^2(n_{10}+n_{11})^2},$$

$$\hat{C}ov(\hat{\tau}_2,\hat{\upsilon}_1) = -\frac{s_{01}r_{00}(r_{01}+r_{11}) + r_{01}\left[s_{11}(s_{00}+s_{01}) + s_{01}(r_{01}+r_{11}+s_{00}+s_{01})\right]}{(n_{00}+n_{01})^2(n_{01}+n_{11})^2},$$

$$\hat{C}ov(\hat{\upsilon}_1,\hat{\upsilon}_2) = \frac{s_{10}r_{00}(s_{00}+s_{01}) + s_{00}\left[r_{00}^2 + r_{01}r_{10} + r_{00}(r_{01}+r_{10}+s_{00}+s_{01})\right]}{(n_{00}+n_{01})^2(n_{00}+n_{10})^2},$$

$$\hat{C}ov(\hat{\tau}_1,\hat{\upsilon}_1) = 0 \text{ and } Cov(\hat{\tau}_2,\hat{\upsilon}_2) = 0.$$

## Appendix II

Wu [15] has studied the simultaneous comparison of the *PVs* solving the global hypothesis test (1) proposing an extension of the McNemar test, and the test statistic is

$$\chi^2 = \frac{(s_{10}-s_{01})^2}{s_{10}+s_{01}} + \frac{(r_{10}-r_{01})^2}{r_{10}+r_{01}},$$

whose distribution is approximately a chi-squared with two degrees of freedom when the sample size is large. Wu demonstrated through simulations that this test statistic shows good asymptotic behavior both in terms of type I error rate and power, and its power is also equal to or better than that of the Roldán-Nofuentes method [14].

## Supplementary Materials

*RMCPV*: a function in *R* to solve the global hypothesis test when a binary covariate is observed in all individuals

*RMCPVDC*: a function in *R* the global hypothesis test when a dummy covariate is created.

## Disclosure statement

No potential conflict of interest was reported by the authors.

## Funding information

This work was supported by Grant "PID2021-126095NB-100" funded by MCIN/AEI/10.13039/501100011033 and by "ERDF, EU".

## Acknowledgments

We thank the three referees, the Associate Editor and the Editor of Journal of Biopharmaceutical Statistics for their helpful comments that improved the quality of the paper.